\documentclass[iop,apj]{emulateapj}
\usepackage{apjfonts}
\usepackage{color}
\usepackage{subfigure}
\usepackage{xspace}

\newcommand{\unitspace}{\ensuremath{\;}}
\newcommand{\usp}{\unitspace}
\newcommand{\numberspace}{\ensuremath{\;}}

\newcommand{\unitstyle}[1]{\ensuremath{\mathrm{#1}}}
\newcommand{\power}[2]{\ensuremath{{#1}^{#2}}}

\newcommand{\centi}{\unitstyle{c}}
\newcommand{\kilo}{\unitstyle{k}}

\newcommand{\meter}{\unitstyle{m}}

\newcommand{\second}{\unitstyle{s}}

\newcommand{\Kelvin}{\unitstyle{K}}
\newcommand{\K}{\Kelvin}  

\newcommand{\cm}{\centi\meter}
\newcommand{\gram}{\unitstyle{g}}

\newcommand{\grampercc}{\gram\usp\power{\cm}{-3}} 

\newcommand{\erg}{\unitstyle{ergs}} 

\newcommand{\km}{\kilo\meter}   
\newcommand{\kms}{\km\usp\power{\second}{-1}} 
\newcommand{\cms}{\cm\usp\power{\second}{-1}} 

\newcommand{\Msun}{\ensuremath{M_\odot}}

\newcommand{\yr}{\unitstyle{yr}}        

\newcommand{\unit}[2]{\ensuremath{#1\numberspace\mathrm{#2}}}

\providecommand*{\eg}{\emph{e.\,g.}\xspace}%
\providecommand*{\ie}{\emph{i.\,e.}\xspace}%

\newcommand{\pder}[2]{\ensuremath{\frac{\partial#1}{\partial#2}}}
\newcommand{\pders}[2]{\ensuremath{\frac{\partial^2#1}{\partial{#2}^2}}}

\newcommand{\ee}[1]{\ensuremath{\times \power{10}{#1}}}

\newcommand{\eq}[1]{Equation (#1)}
\newcommand{\eqs}[1]{Equations (#1)}
\newcommand{\eqref}[1]{\eq{\ref{#1}}}

\newcommand{\tabref}[1]{Table~\ref{#1}}
\newcommand{\figref}[1]{Figure~\ref{#1}}
\newcommand{\secref}[1]{Section~\ref{#1}}

\newcommand{\SNeIa}{SNe~Ia\xspace}
\newcommand{\SNIa}{SN~Ia\xspace}

\newcommand{\Mch}{\ensuremath{M_\mathrm{ch}}\xspace}

\newcommand{\slam}{\ensuremath{s_\mathrm{l}}\xspace}
\newcommand{\slamO}{\ensuremath{\slam^0}\xspace}
\newcommand{\dlam}{\ensuremath{\delta_\mathrm{l}}\xspace}
\newcommand{\dlamO}{\ensuremath{\dlam^0}\xspace}
\newcommand{\sturb}{\ensuremath{s_\mathrm{t}}\xspace}
\newcommand{\sturbD}{\ensuremath{s_{\mathrm{t}\Delta}}\xspace}
\newcommand{\turb}{\ensuremath{u^\prime}\xspace}
\newcommand{\turbD}{\ensuremath{\turb_\Delta}\xspace}
\newcommand{\us}{\ensuremath{\frac{\turbD}{\slamO}}\xspace}
\newcommand{\Dd}{\ensuremath{\frac{\Delta}{\dlamO}}\xspace}
\newcommand{\meancurve}{\ensuremath{\left|\left<\divr\bvec{n}\right>_s\right|}\xspace}
\newcommand{\PrN}{\ensuremath{\mathrm{Pr}}\xspace}
\newcommand{\Le}{\ensuremath{\mathrm{Le}}\xspace}
\newcommand{\Ze}{\ensuremath{\mathrm{Ze}}\xspace}

\newcommand{\ReN}{\ensuremath{\mathrm{Re}}\xspace}
\newcommand{\ReD}{\ensuremath{\ReN_\Delta}\xspace}
\newcommand{\gibson}{\ensuremath{\ell_\mathrm{G}}\xspace}
\newcommand{\lt}{\ensuremath{\ell_t}\xspace}

\input{vectors}
\newcommand{\code}[1]{\textsc{#1}\xspace}

\newcommand{\flash}{\code{flash}}

\newcommand{\nuclei}[2]{\ensuremath{\mathrm{^{#1}#2}}\xspace}

\newcommand{\carbon}[1][12]{\nuclei{#1}{C}}

\newcommand{\oxygen}[1][16]{\nuclei{#1}{O}}

\newcommand{\nickel}[1][58]{\nuclei{#1}{Ni}}

\defcitealias{Colietal00}{CVDP}
\defcitealias{Charetal02a}{CMV}
\defcitealias{Schmetal06a}{SNH}
\defcitealias{Schmetal06b}{SNHR}

\shorttitle{Turbulence--Flame Interaction Model for Astrophysical Flames}

\begin{document}

\submitted{Accepted to the Astrophysical Journal February 16, 2014}
\title{Power-Law Wrinkling Turbulence--Flame Interaction Model for Astrophysical Flames}

\author{
Aaron P. Jackson\altaffilmark{1,2},
Dean M. Townsley\altaffilmark{3},
Alan C. Calder\altaffilmark{4,5}
}

\altaffiltext{1}{
National Research Council Research Associateship Program
}
\altaffiltext{2}{
Laboratories for Computational Physics \& Fluid Dynamics,
Naval Research Laboratory, Washington, DC
}
\altaffiltext{3}{
Department of Physics \& Astronomy,
The University of Alabama, Tuscaloosa, AL
}
\altaffiltext{4}{
Department of Physics \& Astronomy,
The State University of New York - Stony Brook, Stony Brook, NY
}
\altaffiltext{5}{
New York Center for Computational Sciences,
The State University of New York - Stony Brook, Stony Brook, NY
}

\begin{abstract}
We extend a model for turbulence--flame interactions (TFI) to consider
astrophysical flames with a particular focus on combustion in type Ia 
supernovae. The inertial range of the turbulent cascade is nearly 
always under-resolved in simulations of astrophysical flows, requiring
the use of a model in order to quantify the effects of subgrid-scale
wrinkling of the flame surface.
We provide implementation details to extend a well-tested 
TFI model to low-Prandtl number flames for use in the compressible 
hydrodynamics code \flash. A local, instantaneous measure of the turbulent 
velocity is calibrated for \flash and verification tests are performed.
Particular care is taken to consider the relation between the subgrid rms
turbulent velocity and the turbulent flame speed, especially for 
high-intensity turbulence where the turbulent flame speed is not expected
to scale with the turbulent velocity. Finally, we explore the impact of 
different TFI models in full-star, three-dimensional simulations of type
Ia supernovae.
\end{abstract}

\keywords{hydrodynamics---nuclear reactions, nucleosynthesis,
abundances---supernovae: general}

\section{Introduction}
\label{sec:intro}

Type Ia supernovae (SNe) are bright stellar explosions that
are characterized by strong P Cygni features in Si and by a
lack of hydrogen in their spectra.  It has generally been accepted
that these events follow from the thermonuclear incineration of
a degenerate stellar core known as a white dwarf (WD) that
produces $\sim \unit{0.6}{\Msun}$ of
radioactive \nickel[56], the decay of which powers the light curve
(see~\citealt{filippenko_1997_aa,hillebrandt_2000_aa,Ropke06,Caldetal12}, and
references therein for an overview), and direct evidence for this paradigm
was recently established by observations of 
SN 2011fe \citet{nugentetal2011,brownetal2012} and \citet{bloometal2012}. 
The light curves of \SNeIa have the
property that the brightness of an event is correlated with its duration.
This ``brighter is broader'' relation~\citep{Phil93} is the basis for
light curve calibration that allows use of these events as distance
indicators for cosmological studies (see~\citealt{Conletal11} for a
contemporary example).

While there is agreement on the general properties of thermonuclear SN,
the progenitors of these events are not definitively known and their
determination is the subject of active research.  The possible progenitor
systems are generally divided into two categories: single-degenerate
(SD;~\citealt{WhelIben73,Nomo82_1,iben_1984_aa}) and double-degenerate
(DD;~\citealt{iben_1984_aa,webbink_1984_aa}) (\ie, a binary system
composed of either one WD or two).  In the SD scenario, a
\carbon-\oxygen WD accretes material from a main sequence
or red giant companion. Either the WD accretes material from its companion
until it approaches the Chandrasekhar mass (\Mch) sufficiently heating the
core to fuse C and begin the thermonuclear runaway~\citep{Nomo82_1},
in which core convection eventually leads to flame ignition~\citep{Woosetal04},
or a layer of He detonates on the surface of a sub-\Mch WD driving
a compression wave into the core sufficiently strong to trigger a
second detonation (``double-detonation'';~\citealt{Livn90}). In the DD
scenario, two WD's merge via gravitational radiation within a Hubble
time. The less massive WD will be tidally disrupted and accreted onto
the primary. As long as the accretion rate is not too high to ignite
C at the edge~\citep{saio_1985_aa,saio_1998_aa, saio_2004_aa}, it is
possible for the primary to gain enough mass to approach \Mch and explode
as a \SNIa via the same explosion mechanism as the \Mch-WD in the SD
channel~\citep{yoon_2007_aa}. More recent studies, however, modeling 
the accretion of an alpha disk driven by the magnetorotational 
instability find significantly higher accretion rates \citep{schwabetal2012,jietal2013}. Additionally, \citet{Pakmetal10,Pakmetal11}
recently showed that under certain conditions, the merger is violent
enough such that C ignition at the edge may launch a detonation into the
primary WD leading to a \SNeIa with the brightness primarily determined by
the mass of the primary. More recent studies by \citet{danetal2011, danetal2012,
raskinetal2012} find rather different ranges of system masses for successful
``violent'' detonations.

Chandrasekhar-mass progenitor models have been widely studied as a possible
origin of \SNeIa, although the debate between the SD
vs.\ DD channel is still very active. One-dimensional delayed detonation
models have successfully reproduced many observed features that
agree with ``normal'' SNe Ia~\citep{HoefKhok96}.
Chandrasekhar mass detonation-only
models have been ruled out due to the over-production of
\nickel[56]~\citep{Nomo82_2,Nomoetal84} and deflagration-only models
cannot account for brighter \SNeIa~\citep{ropke_2007_ac}. A delayed
detonation mechanism best agrees with many observations of multi-band
light curve shapes, nucleosynthetic yields, spectral evolution, and SN
remnants~\citep{HoefKhok96,Badeetal03, Kase06,KaseWoos07,Woosetal07}.
This mechanism implies a phase of subsonic burning in which the star
may respond to thermonuclear burning and expand before a detonation,
or supersonic reaction wave, is somehow initiated to burn any remaining
fuel on timescales much shorter than the dynamical timescale of the
explosion.  The standard delayed-detonation model is the so-called
deflagration-to-detonation transition (DDT) scenario~\citep{Khok91a}
and variations on the DDT scenario, including pulsational 
detonations~\citep{arnettlivne94a,arnettlivne94b,HoefKhok96,
BravGarc06}, gravitationally-confined 
detonation~\citep{PlewCaldLamb04,Jordan2008Three-Dimension}
\citep[see also][]{Seitetal09det,chamulaketal2012,jordanetal2012a}, 
and a host of models 
in which a deflagration transitions to a detonation when
the right local conditions are met \citep{1986SvAL, woosley90,
hoflich.khokhlov.ea:delayed, HoefKhok96, KhokOranWhee97, NiemWoos97, hwt98, 
Niem99} have also been explored.

Thermonuclear flames for compositions, densities, and temperatures
characteristic of the C-O WD near conditions for explosive C-burning
are spatially thin owing to the extreme temperature dependence of the
driving reaction, $\carbon+\carbon$~\citep{timmes_1992_aa}.  A common
computational challenge in modeling astrophysical explosions is that
typically the physics of combustion is characterized on length scales
well below that accessible in simulations that resolve the macroscopic
explosive event. In the context of \SNeIa, combustion initially proceeds
subsonically, driven by the nuclear fusion of carbon. The flame width
of a laminar carbon flame at the densities relevant in the WD is
$\lesssim \unit{\power{10}{-1}}{\cm}$, while the WD diameter is $\sim
\unit{\power{10}{8}}{\cm}$. The range of length scales involved in the
calculation necessitates the use of a model flame that is resolvable by
the computational domain.

Within the standard Chandrasekhar-mass models of \SNeIa, the rising
temperature in the core allows carbon fusion before collapse. However,
the energy generated from carbon fusion is carried away by
convection and a thermonuclear flame is not born for yet another
$\sim\unit{\power{10}{3}}{\yr}$. By the time the energy generation rate
exceeds the combined cooling rate from convection and free-streaming
neutrinos, the convective region within the core of the WD encompasses
$\sim\unit{0.8-1.0}{\Msun}$ with a root-mean-squared velocity $v_{\rm
rms} \sim \unit{300}{\kms}$~\citep{Zingetal09}. While the character of
this flow is still the subject of active research, this $v_{\rm rms}$ is
higher than the propagation speed of the deflagration front and therefore
the turbulence generated
by convection is expected to influence the subsequent thermonuclear flame.
The magnitude of the effect on the deflagration phase and the subsequent
nucleosynthetic yield has not been fully explored in full \SNIa\ simulations,
but the result is that the turbulence-flame interaction (TFI) is 
essential from the time of first ignition of the deflagration.

The fundamental nature of TFI and, as a result, the best methods to model
it in necessarily under-resolved large eddy simulations (LES) is still an
outstanding problem in combustion \citep{Dris08}.
Thus, there are presently a variety proposed methods for handling LES TFI
that draw on different assumptions and characterizations of the process
and rely on different numerical techniques. Due to the prohibitive 
computational costs involved in validating these assumptions with direct
numerical simulation (DNS) for conditions relevant in the \SNIa, one path
forward is to address these uncertainties directly by exploring
a variety of assumptions for TFI and evaluating the sensitivity to these
assumptions. This approach allows effort to be focused on the features of TFI
modeling that are most salient to the \SNeIa problem.  As will be discussed in
section \secref{sec:background:snia}, the TFI model presented here is
intended to complement previous work on TFI in \SNeIa, most notably by
\citet{Schmetal06a,Schmetal06b}, by varying some
assumptions about the propagation of turbulence on unresolved scales and
some other technical details.

In \secref{sec:background}, we provide and overview of turbulence and flames,
and we briefly discuss differences in current
flame modeling approaches.  In \secref{sec:flash} we briefly review the
important aspects of the \flash hydrodynamics code that we use to perform
simulations.  In \secref{sec:turbulence}, we describe the
implementation details of the method used to measure resolved turbulence
following \citet[hereafter \citetalias{Colietal00}]{Colietal00} and
calibrate the operator to the \flash code.  In \secref{sec:tfimodel}, we
describe the assumptions made to implement a subgrid scale TFI model
developed by \citet[hereafter \citetalias{Charetal02a}]{Charetal02a} that
follows a subgrid flame surface density approach. The TFI model provides
a turbulent front propagation speed given the local turbulent intensity and
laminar flame properties.
In \secref{sec:verification}, we provide simple test problems to verify
our scheme behaves as expected.
As an aside in \secref{sec:future}, we provide
direction for expanding the model to new regimes of validity in future
works.
 We then compare three different TFI models in
full-star, three-dimensional (3D) simulations in \secref{sec:sims} to
highlight differences in model assumptions.  We summarize our findings 
and conclude in \secref{sec:conclusions}.

\section{Background on Turbulence and Flames}
\label{sec:background}

We begin by outlining some general concepts of turbulence and flames  
in order to frame our discussion of model flames and TFI.
With this introduction and discussion of the important physical length 
scales and characteristics in the supernova, we review briefly what 
techniques are available and in use to treat flames in simulations of 
\SNeIa.  This discussion motivates the current study by setting out 
the issues that we hope to address in our numerical models with the 
TFI presented in this work. Our discussion will focus
mainly around techniques used previously in the astrophysical literature
and those important to support the methods presented here.  For overviews
of flame techniques in use in the combustion literature, the reader is
referred to \citet{Dris08} and \citet{PoinVeyn05}.

\subsection{Turbulence}

Turbulence occurs in fluid flow under conditions in which the length 
scales of the flow are significantly greater than the scale of the
viscous dissipation of energy, which allows for disordered, stochastic 
motion and the cascade of energy from large scales to small. Turbulent 
flows are common in Nature, and influence related phenomena such as mixing 
and combustion. Turbulence is incompletely understood and research
into the nature and properties of turbulent flow continues. 

The contemporary understanding of the theory of turbulent 
flow follows from the work of~\citet{Kolmogorov41,Kolmogorov91}, who 
introduced the idea of a self-similar cascade of energy from large
scales to small. The range between the integral or large driving scale 
of the flow and the viscous dissipation scale is known as the inertial 
regime, and depends on the Reynolds number, 
$\ReN$, 
the ratio of inertial forces to
viscous forces, which we may define as $\ReN = \lt \turb / \nu$, where
$\lt$ is the integral scale, $\turb$ is the turbulent intensity,
and  $\nu$ is the viscosity. A high Reynolds number implies a wide 
inertial range and turbulent flow. The turbulent intensity measures
the relative strength of the turbulence and is defined as
$\turb \equiv v'_t / \bar{v}$, where $v'_t$
is the root-mean-square of the turbulent velocity fluctuations and $\bar{v}$
is the mean velocity.  

The self-similar nature of turbulence in the Kolmogorov model implies
that the turbulent flow demonstrates the same statistical properties on 
many scales, which allows the description of properties of the turbulent 
flow via scaling laws. The case of incompressible Kolmogorov turbulence 
is the famous example of a scaling law \citep{clvmhd2003}. If $v_l$ is 
the velocity at scale $l$ within the inertial range, the kinetic 
energy $\sim v^2_l$ is transferred to the next scale down in one 
eddy turnover time, $t_{\rm cascade}=l/v_l$. Kolmogorov theory thus holds that the
energy transfer rate is scale-invariant
\begin{equation} 
\frac{v_l^2}{t_{\rm cascade}} = \mbox{constant} \; , 
\end{equation} 
which yields the Kolmogorov scaling
\begin{equation} 
v_l \propto l^{1/3} \; , 
\end{equation} 
which in turn implies the -5/3 power law for the power
spectrum
\begin{equation} 
E\left(k\right) \propto k^{-5/3} \; .
\end{equation} 
Another example of a scaling law we apply below for the case
of turbulent combustion is Damk\"ohler scaling, 
which relates the average flame speed
to the turbulence intensity, $s_t\approx \turb$.

Experimental and theoretical work, however, indicates
that Kolmogorov's theory of turbulence is incomplete because 
self-similarity is not observed due to the fact that dissipation within 
the fluid is intermittent in both space and time rather than homogeneous. 
This situation suggests
other scaling laws are necessary for a more accurate description of 
turbulent flow 
\citep[See][for more thorough discussion and an example]{Fishetal08}.

\subsection{Flames}

We use the term flame or deflagration front to indicate a localized
region of exothermic reaction that is propagated into the unreacted
material by the diffusion of heat.  Assumption of locality means that
there exists a well-defined separation between unreacted fuel and fully-reacted
products. We note, however, that the diffusion of heat implies a
preheating region that extends ahead of the reaction zone, so the
assumption of locality applies only to the region of the reaction
and not the temperature field.
This assumption makes it conceptually
similar to a surface that divides space into these two types of regions,
though this surface is assumed to have some inherent thickness associated
with the detailed mechanism of its reaction and propagation. An important
parameter for characterizing astrophysical flames is the Lewis number,
$\mathrm{Le} = \frac{D_{\mathrm{th}}}{D_{s}}$ where $D_{\mathrm{th}}$ is
the thermal diffusivity and $D_{s}$ is the material diffusivity.
In the case of astrophysical flames, the Lewis number is asymptotically 
large, \ie, thermal diffusion dominates over species diffusion.

\subsection{Length Scales}

\begin{figure}[tb]
\centering
\includegraphics[width=0.6\columnwidth]{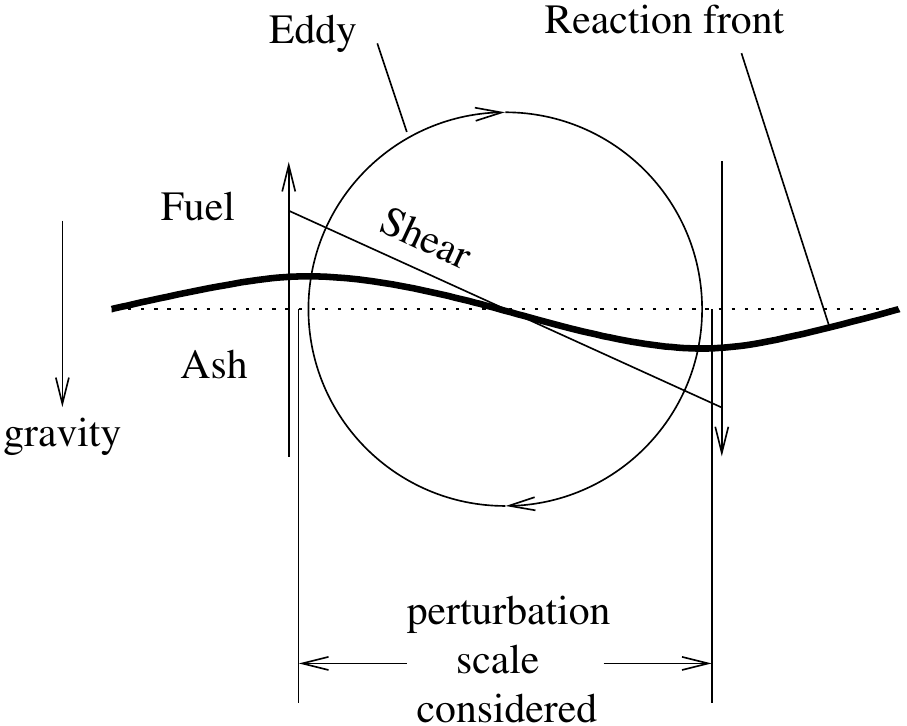}\\
\vspace{1em}
\includegraphics[width=\columnwidth]{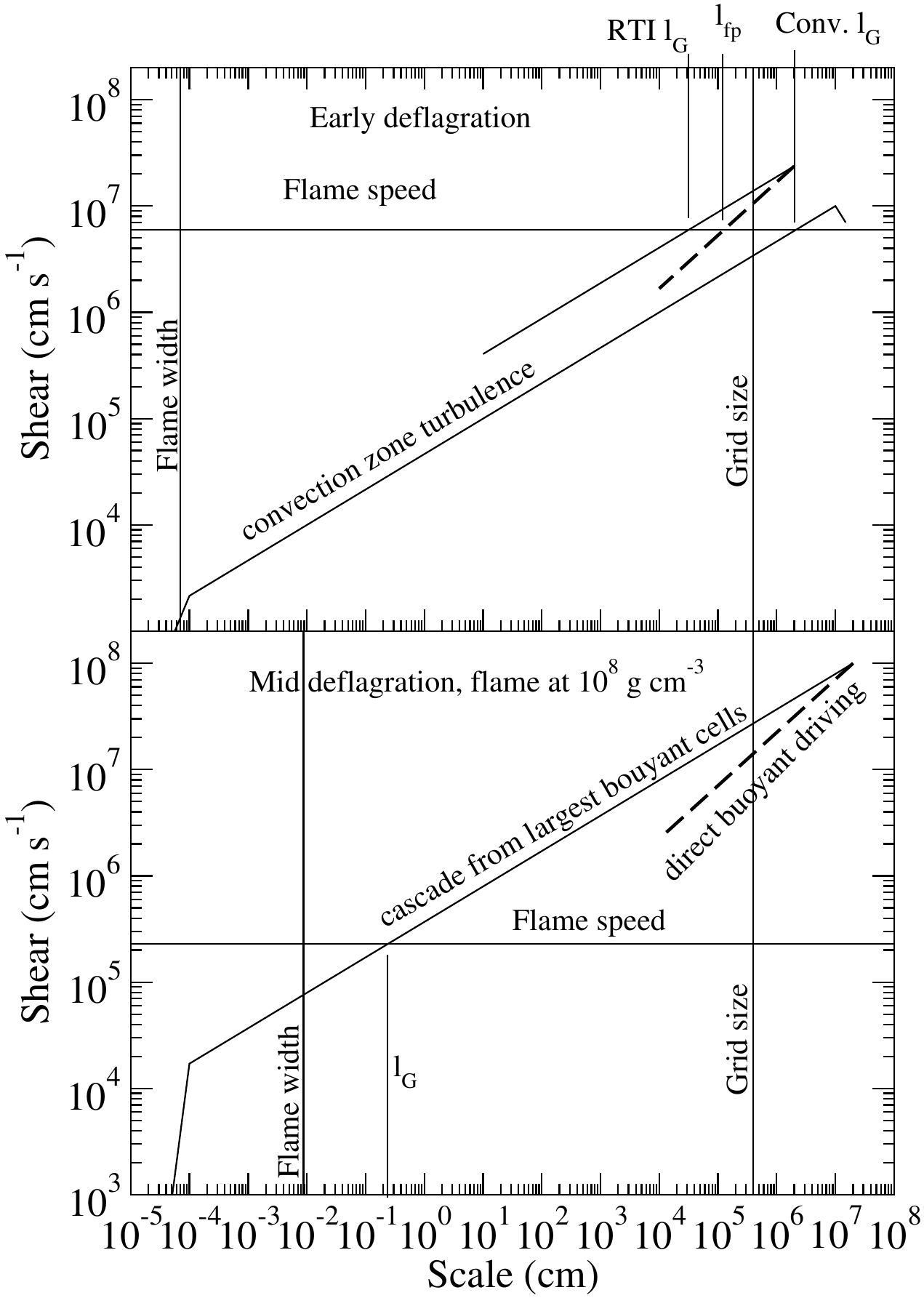}
\caption{
Top Diagram: Competing shearing contributions to wrinkling of flame front on
some perturbation length scale.
Bottom Plots: Shear strength from turbulence or buoyancy for perturbations to
the flame surface on various scales two times during the deflagration, at
early times (top) when $\rho\sim 2\times10^9$~g~cm$^{-3}$, and moderate times
when $\rho\sim 10^8$~g~cm$^{-3}$.  Solid lines indicate shear due to eddies
in the turbulence cascade, while dashed lines indicate strength of buoyancy
driving of moderate amplitude flame surface perturbations.  Both of these
vary with scale in the way shown.  At early times two turbulence cascades are
shown, the lower being the turbulence from the pre-deflagration convection
zone, and the higher being due to the self-generated turbulence of the rising
burned regions.
\label{fig:scales}
}
\end{figure}

Discussion of turbulence, flames, and turbulence-flame interactions
necessitates discussion of the length scales of the problem in question,
which we briefly outline here. This overview describes the length scales of
physics and physical effects that are most relevant to our approach to the
problem of premixed flames and turbulence-flame interactions in type Ia
supernova explosions and is therefore incomplete. \figref{fig:scales}
illustrates how the the flame might be wrinkled by shear on various scales,
and how this is related to some physical and computational scales in our
problem.  The diagram at the top of \figref{fig:scales} demonstrates the two
principal competing wrinkling mechanisms that will be considered for
arbitrary scale.  An eddy in the fluid may wrinkle the flame by simply
advecting adjacent parts of the front in opposite directions.  Buoyancy will
cause any perturbation of the type shown to grow in size due to the
Rayleigh-Taylor instability (RTI).  We wish to compare the strength of the
shear given by each of these two processes with each other and with the flame
propagation speed at a variety of scales and at different densities.

The flame is born in the core of the white dwarf where 
$\rho \approx 2\times\unit{\power{10}{9}}{\grampercc}$. At this high
density, the flame is well-defined with a width $\delta_l \sim \unit{10^{-4}}{\cm}$. 
As the flame propagates into lower density material, the width increases
and the flame become less well-defined, with disparity between the 
rate of C-fusion and the rate of other constituent reactions and the 
effect of turbulence becoming increasingly
important~\citep{timmes_1992_aa,Caldetal07,ZingDurs07,Aspdetal08b,orvedahletal2013}. 
The flame width and speed at densities of $2\times 10^9$~g~cm$^{-3}$ and
$10^8$~g~cm$^{-3}$ are shown by the left vertical line in the top and bottom
plots respectively in \figref{fig:scales}.  The corresponding flame
propagation speeds are shown by the horizontal lines in each.

The shearing characteristics of turbulence can be estimated from
from the approximate driving and the cascade.  The convection zone is thought
to be driven on scales $\sim10^7$~cm, or perhaps somewhat less, with rms
speeds on that scale of order $10^7$~cm~s$^{-1}$
\citep{Woosetal09,Zingetal09}.  From this outer scale, the cascade will
proceed toward smaller eddies, giving $v'\propto l^{1/3}$, the standard
turbulence cascade, where $v'$ can be used as a measure of the shear on a
characteristic scale $l$.

The flame is subject to the effects of fluid instabilities
and turbulence, both of which wrinkle the flame and thereby boost 
burning~\citep{markstein1964}.  The subsonic burning
front that begins near the center of a massive white dwarf is subject
to Kelvin-Helmholtz, Landau-Darrieus, and Rayleigh-Taylor instabilities
\citep{KhokOranWhee97, khokhlov+00, hillebrandt_2000_aa}.
The principal fluid instability influencing the flame is the Rayleigh--Taylor 
(RT) instability~\citep{rayleigh1882,taylor+50,chandra+81}, occurring 
because the gradient of density across the flame is opposite the 
direction of acceleration of gravity.
The evolution of the flame is also influenced by the Kelvin--Helmholtz (KH)
shear instability~\citep{helmholtz1868b,helmholtz1868a,kelvin1871} and the
Landau--Darrieus
instability~\citep{darrieus:1938,landau1944,pelce1988,landau.lifshitz:fluid,bychkovliberman1995,ropke_2004_aa}
on scales similar to or smaller than those of the RT instability. 

The shear contribution to flame wrinkling due to RTI on some arbitrary scale
$l$ can be estimated by considering a perturbation like that shown in the top
diagram in \figref{fig:scales}.  Ignoring curvature, the exponential growth
rate of such a perturbation is $r=\sqrt{g_{\rm eff}/l}$, where $g_{\rm eff}$
is the reduced gravity.  For a modest perturbation, $\sim 0.1l$, like that
shown in the figure, this can be converted to an effective shearing speed.
$v'\sim lr=\sqrt{g_{\rm eff}l}$.  This allows a simple estimate of how the
shearing (wrinkling) strength of the RTI depends on scale and allows an
estimate to be made based on a typical plume size and gravity.  This is shown
by the thick dashed lines in \figref{fig:scales} for the early and middle
deflagration epochs shown.  In the early epochs, the plumes are expected to
be only a few$\times 10^6$~cm across, while later they may reach sizes of
several$\times10^7$~cm or more as they rise into the outer parts of the WD,
whose radius is about $2\times 10^8$~cm.

The RTI will also act as a driver of turbulence on all scales on which it is
active.  This turbulence will cascade to smaller scales via eddy break up.
This buoyancy-generated turbulence is indicated in \figref{fig:scales} by the
upper thin solid line in the upper plot and the single thin solid line in the
lower plot.  The shear due to the cascade from larger scales is generally
stronger than the driving available from the RTI on that same scale.  Note
however, that this is a steady-state view, and we are neglecting how the
transient behavior through which this develops might behave.

Flame wrinkling is expected to cut off when $v'\lesssim s_l$, the laminar
flame speed, as the flame
propagation will smooth out any induced wrinkles.  For turbulence this is
called the Gibson
scale~\citep{peters1988,niemwoos1997,peters2000,roepkeniemhill2003,Aspdetal08b},
the size of the smallest eddy, with a turnover velocity $\gtrsim$ laminar
flame speed.  With a turbulence cascade that gives $v'(l)=V(l/L)^{1/3}$, this
gives a Gibson scale of
\begin{equation}
l_G = L\left(\frac{s_l}{V}\right)^3,
\end{equation}
where $V$ is the velocity on some scale $L$, typically taken to be the
driving scale, but can be taken as any scale in the inertial range of the
turbulence.
An analogous scale exists for the the RTI, 
fire-polishing length
\begin{equation}
l_{\rm fp} = 4 \pi \frac{\slam^2}{g_{\rm eff}} \; ,
\end{equation}
where ${g_{\rm eff}}$ is the effective
gravitational acceleration~\citep{timmes_1992_aa,zingale_2005_aa}.  
The smallest scale of interest is the viscous dissipation scale, 
$\eta_K$, which in the case of the degenerate plasma of the
white dwarf is due to electron-ion collisions~\citep{Woosetal09}, and is
indicated by the downturn in the turbulence spectrum.  This downturn is not
included in the cascade from buoyancy at early times since the time-dependence
of this development may be significant.
The Gibson scale and the fire-polishing scale represent the smallest scale of
wrinkling of the flame surface by turbulence and RTI respectively, since in
all cases here they are larger than $\eta_K$.

As can be seen from \figref{fig:scales}, the only time at which any of the
Gibson scale, flame width, or fire polishing scales are resolved in a typical
$\sim 4\times 10^5$~cm resolution supernova simulation is the Gibson scale
for the flame subject to the pre-ignition convection field.  As soon as any
significant RTI takes place, both the Gibson scale and fire polishing scales
are below the scale of the grid.  The model presented here attempts to model
the behavior of the flame between the resolved scale and the smallest
wrinkling scale.  It appears this inner scale will typically be the
Gibson scale created by the cascade from the driving due to the largest
buoyant cells.  Eventually, in the late deflagration, as the flame width
becomes even larger, this wrinkling scale will be smaller than the flame
width, leading to strong distortion of the laminar flame structure and
possibly a deflagration to detonation transition.

In all conditions the main driving of the turbulence being modeled,
pre-existing convection, RTI, and large K-H scales related to the RTI plumes,
are on scales far enough above the grid scale that the cascade should be able
to be characterized on resolved scales well enough to infer the subgrid
structure of the cascade.  One possible exception is Kelvin-Helmholtz effects
which may also drive turbulence from smaller scales more similar to the flame
surface scale, which is unresolved.  The TFI models discussed here are
intended to model the extension of turbulence to subgrid scales.  Thus as
long as the driving scales are sufficiently larger than the grid scales for
the turbulence cascade to be properly measured, changes in the driving scale
should be accounted for.

\subsection{Flame Models}

\begin{figure}[tb]
\centering
\includegraphics[width=\columnwidth]{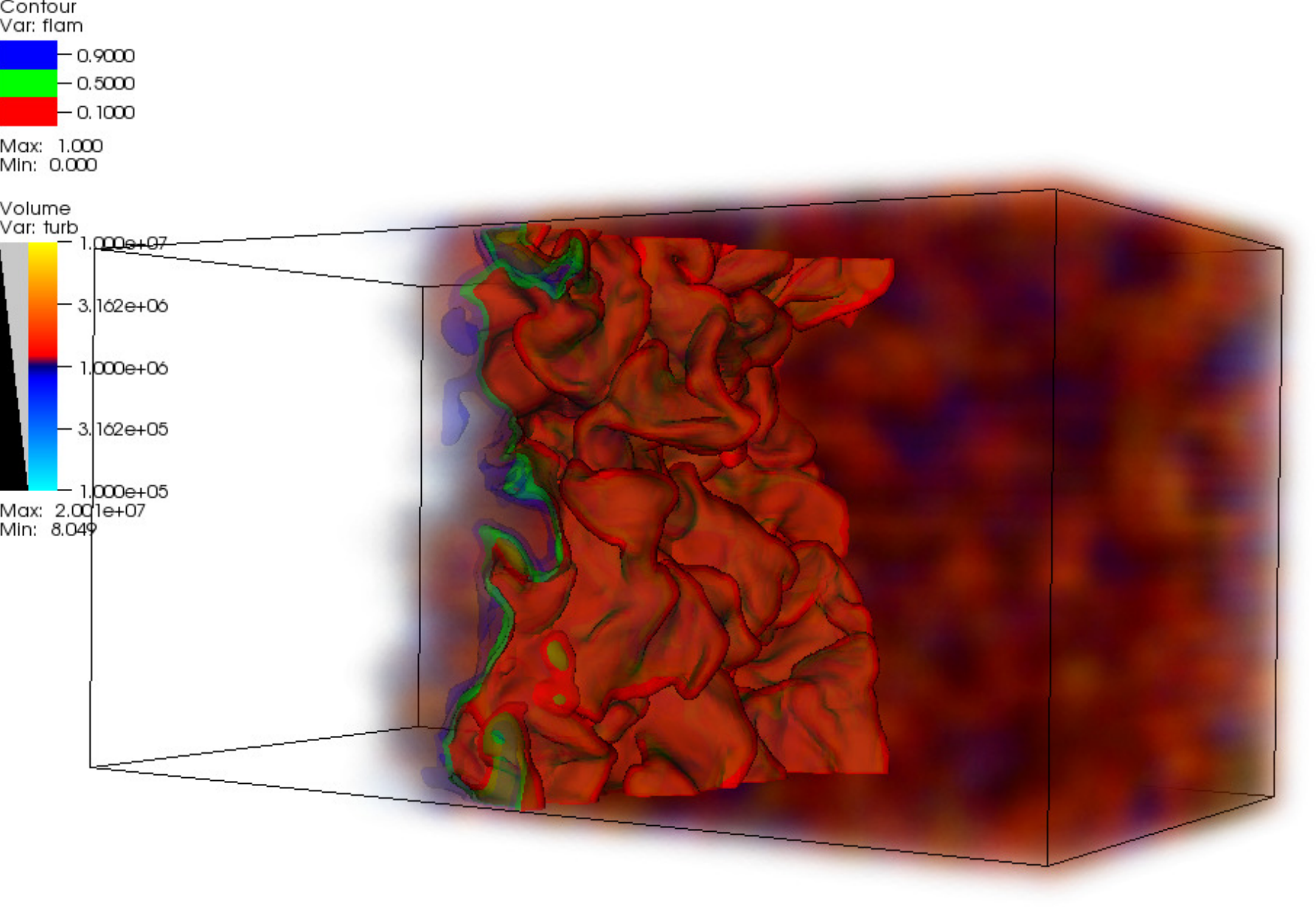}
\caption{
Iso-surface of the reaction progress variable at $\phi=\{0.1,0.5,0.9\}$
(red, green, and blue respectively)
for a reaction-diffusion model flame.
The volume rendering indicates the turbulent strength with stronger
turbulence in orange and weaker in blue. An initially laminar flame
propagating at $\unit{10}{\kms}$ interacts with decaying turbulence
with an initial rms velocity of $\unit{100}{\kms}$.  The channel width is
15~km.
\label{fig:overview}}
\end{figure}

It is convenient to introduce two immediate abstractions that are closely
related to the flame modeling techniques that will be discussed shortly.
One is to define some form of reaction progress variable. We will use
$\phi$, which varies, for example, from 0 to 1 from the fuel to the ash.
The flame is then defined by a field $\phi$ at every point in space. This
definition
abstracts away much of the detail of the flame structure, including the
distinction between its thermal and compositional structure, though $\phi$
can be taken to represent some aspect of one of these two.  A second layer
of abstraction can be introduced (conceptually or explicitly) by
representing the flame as an isosurface, say $\phi=1/2$.  The
(thick-)surface-like region can be highly wrinkled by fluid advection, as
shown in \figref{fig:overview} which shows a model flame (a
reaction-diffusion front) propagating in a turbulent fluid.  The
isosurfaces show regions near the "forward edge" of the flame ($\phi=0.1$)
and the "back edge" ($\phi=0.9$), and demonstrate that the reaction region
as a whole is fairly well characterized by a slightly thickened wrinkled
surface.  What is meant by the thickness of the region and its detailed
structure is not universal, and will depend on what features are being
modeled vs.\ being simulated directly.  Several examples will be given
below.

The turbulence in \SNeIa\ and its relation to and interaction with the
flame provides several challenges. First, the Reynolds number (\ReN)
characteristic of combustion in a degenerate WD is practically 
infinite ($\sim 10^{14}$), which implies that $\lt \gg \eta_k$, where
\lt is the integral scale of the turbulent cascade and $\eta_k$ is the
dissipation scale due to electron-ion collisions~\citep{Woosetal09}.
In fact, even for the weak turbulence present when the flame ignites
($v'_t\sim s_l$, where $v'_t$ is the rms velocity fluctuation on the
integral scale and $s_l$ is the speed of the flame in laminar flow),
$\eta_k$ is typically smaller than the laminar flame width, $\eta_k
<\delta_l$.  This characteristic presents an important distinction from 
laboratory flames, where, for weak turbulence $\eta_k > \delta_l$.  Thus
while in laboratory flames under weak turbulence, there simply are no eddies
active on scales similar to or smaller than the flame thickness. In
astrophysical flames, however, there are always slight perturbations to 
the reaction zone from eddies on these scales.

In addition, the timescale for the explosion is $\sim 1\second$, while the
timescale for turbulence at the integral scale at ignition is comparable
at $\sim 2-20 \second$~\citep{Zingetal09}. The timescale for turbulence
is expected to speed up quickly as driving from the RT KH fluid 
instabilities result from burning cold, dense fuel to hot, light ash in the
deep gravitational well of the degenerate WD. All the while, the star is
expanding due to the release of nuclear binding energy, 
increasing the mean velocity of the fluid elements, which
serves to dampen the turbulent velocity fluctuations in relative terms. All
of these competing processes on the fluid flow occur on similar timescales,
which should not allow one to assume equilibrium turbulence. 
However, due to the necessity to treat unresolved structure in fluid flow
and the lack of availability of an appropriate model, turbulence is
almost always assumed to be in equilibrium on sub-grid scales. This
assumption allows for analytic expressions for the time-averaged
behavior of the unresolved turbulence as a function of resolved, 
larger-scale turbulent fluctuations. Despite the need for a non-equilibrium,
non-Kolmogorov model for unresolved turbulence, development efforts are
still in their infancy, at least in the context of \SNeIa.

For \SNeIa simulations, we are primarily concerned with the effect the
turbulence has on flame propagation.
The modeled flame should respond appropriately to the unresolved modeled
turbulence, in addition to the explicit fluid flow (resolved turbulence)
in the simulation.  Due to their close relationship in implementation, in
a general context the term ``TFI model'' will include both a model of
turbulence and its interaction with the flame, because both must be
specified. However, for this work we will mostly make a distinction
between the turbulence model and the TFI, and use the latter term for the
portion of the model that directly addresses effects on the flame.

While the RT and
KH instabilities play an important role in generating flame surface in
\SNeIa, the inclusion of their combined effects into a self-consistent 
model for under-resolved turbulent combustion is a difficult 
problem~\citep{Khok95,Schmetal06b,Townetal08} that we will not address 
in this work. This choice is
partly due to the desire to assume homogeneous, isotropic turbulence, which
is well described by Kolmogorov turbulence 
theory~\citep{Kolmogorov41,Kolmogorov91}, and
provides a starting point for treating TFI.
The inclusion of either the RT or KH instabilities breaks the assumption
of isotropy, and greatly increases the complexity of the theoretical model;
however, these inclusions may not be necessary provided the simulation
is sufficiently well-resolved~\citep{zingale_2005_ab,Ciaretal09}.

\subsection{Types of Flame Models}

A variety of applicable approaches have been used to model flames
\citep{Dris08,PoinVeyn05}, which we will not attempt to review here in
detail.
Instead we focus the discussion towards the differences
between artificially thickened flames and flame front tracking methods,
because this difference is important for our adaptation of the work in
\citetalias{Colietal00} and \citetalias{Charetal02a} to astrophysical
flames. Artificially thickened flames model the combustion with
similarity to the physical processes, with ``simple chemistry'' using an
Arrhenius law, and thermal and chemical diffusion. However, the reaction
rate and molecular and thermal diffusivities are increased by a thickening
factor, such that the real flame structure is resolved on the computational 
domain but the propagation speed is unchanged~\citep{orourke1979}. On the 
other hand, flame front tracking methods (used in astrophysics) such as the
$G$-equation~\citep{Schmetal06a} and advection-reaction-diffusion (ARD or
ADR) equation~\citep{Khok95}, do not attempt to reproduce physical
features of the flame structure.

The thickened flame approach has the advantage of incorporating various
phenomena naturally, such as effects due to flame curvature and stretch, 
by using the Arrhenius law. This is due to the fact that the basic
physical flame structure is retained,
including the pre-heat zone and the reaction zone. This structure is 
especially important for astrophysical flames in which the Lewis number (\Le)
is nearly infinite, \ie, thermal diffusion dominates over species diffusion.
For large \Le, the local propagation speed depends strongly on curvature, due
to the focusing or de-focusing of heat by thermal conduction.
This approach works well when the thickening factor is not too large,
such that not
much flame structure can hide on unresolved scales; however, for large 
thickening factors, it is not obvious that the thickened flame structure
represents the average behavior of the flame on unresolved scales,
particularly when turbulence is moderately strong. The model flame in this
case will be strained and stretched by the turbulence at the scale of
the computational domain, and the resulting nonlinear burning rate is
likely to be much larger than that expected from the physical flame.

A thickened flame approach would have disadvantages for the
\SNeIa full-star explosion simulations.
For our simulations of \SNeIa, the laminar flame width is
under-resolved by a factor ranging $10^6-10^{10}$. In this case, we do
not expect the model flame structure to behave the same way
as the physical flame on these scales. Curvature effects that should not be
important for weak turbulence will be unduly promoted into severe
effects by the artificial thickening of the flame.  In addition, a single
thickening factor would not be appropriate for the entire range of
physical conditions treated during the explosion.  The range of physical
flame widths varies by many orders of magnitude
\citep{ChamBrowTimm07,timmes_1992_aa}.
For small thickening factors, it is advantageous to have some of the
physical features of flames, such as curvature effects, directly
simulated.  For large thickening factors these effects are more
appropriately moved into the model for the flame dynamics where they can
be controlled and introduced when they are physically relevant.

In order to avoid undesirable issues with a thickened flame, it is useful
to substitute a relatively artificial structure for the reaction front;
preferably one
that does not respond strongly to strain and stretch.
This is necessarily a coarse structure, since it must be resolved in the
simulation.  We use an ADR equation to define a propagating reaction front
that represents the flame front, with the reaction and diffusion terms
determined dynamically to provide a desired front-propagation speed and
interface
width~\citep[see][]{Khok95,VladWeirRyzh06,Townetal07,Townetal09}.  Rather
than having a peaked energy-generation rate in the reaction zone as with
the Arrhenius law, the energy generated from combustion is smoothly
distributed over the interface. This approach has the advantage of having
been demonstrated to produce an acoustically quiet flame front
\citep{Townetal07}.
Additionally, the propagation is effectively due to species diffusion
rather than thermal diffusion; however, it is important to note that the
ADR scheme is not intended to represent physical flame structure.
Another alternative, based on methods widely applied in the combustion
literature and applications is that of the $G$-equation \citep[also called
``level-set'']{reinecke.hillebrandt.ea:new,ropke_2003_aa}, in which the
interface surface between fuel and ash is reconstructed explicitly on the
grid and then propagated using the 
advection of an appropriately defined scalar field.
This case has the advantage of being effectively as thin as possible,
presenting a challenge of what to do at the interface
\citep{ropke_2003_aa}, but still is limited to representing a coarse surface
structure. For the level-set approach as implemented by
\citet{reinecke.hillebrandt.ea:new}, the artificial front is known to produce
spurious velocity oscillations due to the approximations adopted in the 
``passive'' implementation defined in that work and used in astrophysical
simulations.

In any case, a model is required to inform the artificial reaction
front how fast to propagate given local conditions such as the local
thermodynamic quantities, composition, and turbulent intensity. Each approach
has its own advantages and disadvantages as summarized above. We choose to
employ the ADR scheme to both minimize numerical noise and minimize curvature
and strain effects compared to a thickened flame.
Utilizing an acoustically quiet model flame is important in the
context of TFI models to minimize the possibility of providing feedback
into the turbulence measure. While such feedback effects may be possible,
they should be well-controlled and described in the TFI model.

\subsection{TFI Models Used for \SNeIa}
\label{sec:background:snia}

In this work we describe a method to enhance
the laminar front propagation speed of our model flame. This method is
composed of two parts: measuring the subgrid scale (SGS) turbulent intensity
(\secref{sec:turbulence}) and estimating the turbulent flame speed from
the turbulent intensity (\secref{sec:tfimodel}). We utilize an
instantaneous, local measure of the SGS turbulent intensity from the
resolved fluid flow and a TFI model that accounts for inefficient
wrinkling of the flame which will occur at low densities.  As a prelude, this
section provides some context of how our choices relate to and are motivated
by previous work on TFI for \SNeIa.

Treatment of flames in multidimensional simulations of \SNeIa have
followed basically two lines of development, beginning generally from work
of \cite{NiemHill95} on one hand and \citet{Khok95} on the other.  From
the standpoint of TFI, these can be grossly characterized as explicit and
implicit respectively, though the latter does have an explicit model for
RT effects.  Our use of "implicit" is in the sense of ILES (implicit large
eddy simulation) of fluids, in which viscous dissipation is left to
numerical effects (finite resolution) without inserting an explicit
numerical viscosity from a defined model of SGS turbulence \citep[see
e.g.][]{Aspdetal09}.  In this way ``implicit'' means that salient
features are assumed, for at least some portions of parameter space, to
transfer to the grid scale in a way which does not change the gross
simulation outcome.  We will not discuss implicit methods in any more
detail here, as this work is concerned with the development and
exploration of explicit TFI, for which we have specific concerns, outlined
below, in relation to previous work on that topic.

Continuing work outlined by \cite{NiemHill95},
\citet[hereafter \citetalias{Schmetal06a}]{Schmetal06a} recently developed
an SGS model to account for under-resolved TFI and applied
it to \SNeIa with the inclusion of under-resolved RT modes in
\citet[hereafter \citetalias{Schmetal06b}]{Schmetal06b}. Part of the
motivation for the present study is out of interest in understanding
the impact of some of the modeling choices made in
\citetalias{Schmetal06a} on the results presented in
\citetalias{Schmetal06b} and subsequent work \citep[e.g.][]{Seitetal11}.
\citetalias{Schmetal06a} developed a detailed model of SGS turbulence,
which separately treats the creation, transport and dissipation of SGS
turbulent energy. While this model is a major advancement in the
treatment of TFI as it allows for the quantification of SGS turbulence,
several features of the model are concerning in the context of \SNeIa that
must be discussed.

First, we outline our concern with the posited diffusive term in the
\citetalias{Schmetal06a} SGS turbulence model.
In the process of calibrating their closure relations,
\citetalias{Schmetal06a} demonstrated that computing the diffusion of SGS
turbulent energy via gradient-diffusion yields an incorrect estimate of
the direction of diffusive flux.
When directionality is considered using a dot product between the
gradient direction and the computed flux of SGS turbulent energy, as in
\citetalias[\eq{77}]{Schmetal06a}, the resulting calibrated closure
parameter underestimates the magnitude of diffusive flux by an order
of magnitude. On the other hand, when directionality is ignored, as in
\citetalias[\eq{78}]{Schmetal06a}, the resulting calibration demonstrates
that the magnitude of the gradient in SGS turbulent energy is an 
excellent predictor of the magnitude of the actual flux.
Together these imply that, on average, only $\sim 10\%$ of the flux of SGS
turbulent energy is directed in the gradient direction.
There is no clear way to go about finding the direction of the diffusive flux
from coarse quantities (but see \citetalias{Schmetal06a} for
references to some work in this direction.)
Thus, in order to use the gradient-diffusion method,
\citetalias{Schmetal06a} must choose how much flux to send in the gradient
direction.  They choose to set the closure parameter---the coefficient of
the diffusive flux term---to the value which gives the correct magnitude
of the SGS turbulent energy flux, resulting in a turbulent kinetic Prandtl number of
order 10 (extremely diffusive).  However, inverting the implications
discussed above, this overestimates the flux in the gradient direction by
an order of magnitude.

Our concern for transport of SGS turbulent energy in the gradient direction is
of particular interest for the \SNIa problem.  From
\citetalias[Figures 4--6]{Schmetal06b}, the magnitude of SGS turbulent
energy moved by diffusion is comparable to production, and production occurs
mostly in the ash behind the flame. Our concern is that the direction
of gradient-diffusion is computed to generally align with the direction
of flame propagation (since turbulence is generated behind the flame front),
and the spread of the flame may be dominated by turbulence diffusing from
behind the flame to the flame front. This is quite likely to be the
proper physics, at least in certain portions of the flame. However,
according to the tests of gradient-diffusion presented in
\citetalias{Schmetal06a}, the magnitude of this effect may be
over predicted by a factor of $10$ in their SGS model.

Finally, we must point out that \citetalias{Schmetal06b} utilize an
implementation of the level-set method to describe their model flame, 
which is known to produce oscillations in the velocity field near the
flame front~\citep{reinecke.hillebrandt.ea:new}, and was demonstrated to have
poor stability of the flame surface~\citep{ropke_2003_aa}.
It is not known whether the turbulence
generated by this noise is captured by their SGS turbulence model or how
significant the effect might be.

It is not established that such a detailed model of SGS turbulence is
necessary for treatment of TFI in the \SNIa. The spread of resolved
turbulence across
the grid might be a sufficient quantification of this effect, without
requiring the invocation of a model. As discussed in
\citetalias{Colietal00}, on which our measurement of unresolved turbulence
presented in \secref{sec:turbulence} is based, it is not obvious how
to translate quantities in an SGS turbulence model into an
appropriate "measure" of turbulence that will interact with the flame.
In particular, any determination of turbulence must evaluate to zero
in the limit of a perfectly laminar flame. As a
result, even for simulations which include an SGS turbulence model, 
\citetalias{Colietal00} prefer a separate operator for TFI specifically constructed
to filter out effects due to thermal expansion of the reacted products. (See
\secref{sec:turbulence} for continued discussion.)

The second way in which we attempt to improve treatment of TFI for \SNeIa
is to improve the way in which high turbulence intensities---or
equivalently low laminar flame speeds---are handled.
\citetalias{Schmetal06b} assume the interaction between the
flame front and turbulence is scale invariant~\citep{Poch94}, so that the
flame can wrinkle arbitrarily finely in response to turbulence. However,
for high-intensity turbulent combustion, the turbulent flame speed has
not been observed to scale with the turbulent intensity~\citep{AbdeBrad81}.
The approximation of scale invariance for TFI in degenerate WDs seems
to be appropriate
near the core where the density is high and the ratio of turbulent intensity
to the laminar flame speed is relatively low; however, as the flame
propagates to lower densities and approaches conditions predicted for DDT,
this approximation is no longer valid. This effect has been handled in previous
simulations in an approximate way by inserting a density cutoff in the
flame propagation. The TFI models we are adapting, \citetalias{Colietal00}
and \citetalias{Charetal02a}, are constructed to allow for inefficient
wrinkling, and evaluate this effect based on flame-vortex interaction
efficiencies evaluated with direct numerical simulations (DNS) of unity-\Le
flames. This improvement is less likely to have a direct effect on
explosion outcomes, only becoming important at late times near the DDT,
but may be important for properly localizing the DDT site.  It will also be
important in future work to incorporate results from DNS studies of
flame-turbulence interactions that are specific to the flame in \SNeIa\
\citep{Aspdetal08b,aspdenetal2010,aspdenetal2011} as an improvement over the
more general vortex-flame interactions for simple reactions studied by
\citetalias{Colietal00}.  Improved modeling of flame-vortex interaction in
the later stages of the deflagration may be particularly relevant to
understanding situations in which the DDT might fail
\citep{jordanetal2012b,kromeretal2013}.

\section{Overview of \flash}
\label{sec:flash}

This manuscript concerns the implementation of a TFI model to describe
sub-grid scale processes unresolvable in full-star, 3D simulations of
\SNeIa using the \flash code, version 3. In order to properly implement the model,
\flash is also used to generate turbulence fields and propagate test flames
in simplified geometries. Full details of the numerical code can most
recently be found in \citet{Townetal09} and work referenced therein, but an
overview is provided here for completeness. 
\flash is a Eulerian compressible adaptive-mesh
hydrodynamics code~\citep{Fryxetal00,calder_2002_aa}. \flash uses a
high-order shock-capturing compressible hydrodynamics method, the piecewise
parabolic method \citep[PPM,][]{ColeWood84}, adapted to
treat a general equation of state \citep[EOS,][]{ColeGlaz85}. We use
a tabulated fully-ionized electron-ion plasma
EOS~\citep{timmes.swesty:accuracy,Fryxetal00} appropriate for conditions
in the degenerate core of a massive WD. \flash supports using this
hydrodynamics method on an adaptively refined, tree-structured, non-moving
Eulerian grid. Several
modifications have been incorporated to perform simulations of \SNeIa
including a nuclear burning model, updated flame model with
composition-dependent input laminar flame speeds, and 
specific mesh refinement criteria~\citep[for details, see][]{Caldetal07,
Townetal07,Townetal09,Jacketal10}. While the utilization of adaptive mesh refinement
(AMR) is critical for simulations of \SNeIa, \flash also supports a uniform
grid, which we use for generating turbulence and propagating flames in test
channels.

As motivated above, the unresolved flame is modeled by a resolved
reaction-diffusion front in which a progress variable, $\phi$, varies from 0
to 1 as burning proceeds.  Reaction and diffusion act on the same variable,
so that the dynamics are given by
\begin{equation}
\frac{\partial \phi}{\partial t} + \vec v\cdot\nabla\phi =
\kappa \nabla^2\phi + \frac{1}{\tau}R(\phi),
\end{equation}
where $R(\phi)$ is a reaction function, $\vec v$ is the fluid velocity and
the coefficients $\kappa$ and $\tau$ are chosen along with $R$ to maintain a
resolved reaction front width and the desired spatial propagation speed.
Energy is released linearly with $\phi$ in our test simulations and as
described in \citet{Townetal07} in supernova simulations.  How the speed
should be chosen is addressed in this work as discussed below.  All
components implementing reaction-diffusion and turbulence measurement used in
this work, including test simulations presented in section
\ref{sec:convergence}, will be available in the 4.2 release of the \flash\
Code.

\section{Measuring Unresolved Turbulence}
\label{sec:turbulence}

In this section we outline and calibrate a local differential operator
that will quantify the strength of turbulence at any given region on the
grid.  This will form an input to the TFI model described in
\secref{sec:tfimodel}.  Our ``model'' of SGS turbulence, such as it is,
is simply a Kolmogorov cascade whose local strength at a scale $\Delta$
just above the grid scale $\Delta_x$, is based on the operator
outlined here.  Thus, we implicitly assume that the SGS turbulent velocity
field is homogeneous and isotropic on scales smaller than approximately
$\Delta$ and does not vary significantly in strength on timescales shorter
than the eddy turnover time at the grid scale.
This model, though simple, will be used explicitly in evaluating
how the TFI wrinkles the flame.
As discussed in detail in \secref{sec:background} our model of
turbulence is chosen to complement other approaches such as that of
\citetalias{Schmetal06b}, which instead have a detailed \emph{dynamical}
model for SGS turbulence creation, transport and dissipation.

We want to measure resolved turbulent motions that are not caused by the
expansion of material as it is burned. This component of turbulence can be
argued to exist ahead of a propagating burning front such that it wrinkles
the front and influences the local burning rate. \citetalias{Colietal00}
describe a finite-difference operator ``$OP_2$'' to measure
resolved turbulence with the irrotational velocity component filtered out
\begin{eqnarray}
OP_2\left(\bvec{u}\right) &=& \left( h \Delta_x \right)^3
\left| \curl \left( \lap \bvec{u} \right) \right|
{\rm ,} \label{eq:turb_op} \\
\turbD &=& c^h_2 OP_2 \left( \bvec{u} \right)
{\rm ,} \label{eq:turbD}
\end{eqnarray}
where $u$ is the velocity vector field,
$c^h_2$ is a constant, calibrated below, such that $\turbD$
represents the turbulence on unresolved scales (<$\Delta$), and $h$ is the
index of the stride in the finite difference scheme (an integer, typically
1 or 2).
The irrotational component is filtered out in order to prevent the expansion
inherent in the flame from contributing to the measurement.
The constant $c^h_2$ is determined by requiring
that the kinetic energy of the
turbulent velocity measured be equal to the kinetic energy contained in
the turbulent cascade between $\Delta$ and the physical dissipation scale
($\eta_k$).
The length scale $\Delta$ is a characteristic of the operator,
and therefore depends on $h$ and $\Delta_x$.
Because the dissipation range of resolved turbulent flows
begins well above the grid scale, the length scale $\Delta$ associated
with the turbulence operator typically lies within the dissipation range.
Therefore, $c^h_2$ serves partially
to correct for the effects of numerical dissipation. This formulation
allows some flexibility in $\Delta$ by choosing different values for the
integer stride $h$.  In general, $c^h_2$ will depend on the choice of
$h$ and the numerical method used to evaluate \eqref{eq:turb_op} and the
particular form the numerical dissipation takes~(see
\citealt{Sytietal00} for more details about numerical dissipation with
PPM). We use a fourth-order finite central difference with 
\begin{eqnarray}
\pder{f}{x} &=&
\frac{ f_{i-2h} - 8 f_{i-h} + 8 f_{i+h} - f_{i+2h} }{ 12 h \Delta_x } \\
\pders{f}{x} &=&
\frac{ -f_{i-2h} + 16 f_{i-h} - 30 f_{i} + 16 f_{i+h} - f_{i+2h}}%
{ 12 h^2 \Delta_x^2 } {\rm .}
\end{eqnarray}
For this particular implementation, the length scale associated
with the turbulence operator is approximated by $\Delta = 4 h \Delta_x$,
because the turbulence measured in a particular cell uses velocity
information from cells up to $4 h$ cells away. Other definitions of $\Delta$
could be adopted which would affect the calibration of $c^h_2$ essentially
rescaling the distribution of $OP_2$ to agree with the expected value at
scale $\Delta$.

For \flash, we calibrate $c^h_2$ using a Kolmogorov turbulence cascade
generated by driving fluid motions on large scales.  We drive turbulence
in a triply-periodic Cartesian box with varying resolutions for 1.5
eddy-turnover times ($\tau_e$), where $\tau_e = L / v_{\rm rms}$,
$L = \unit{1.5\ee{6}}{\cm}$ is the size of the box, and $v_{\rm rms}$ is the
root-mean-squared velocity of the resolved flow. We follow the simulation
setup \texttt{StirTurb} distributed with \flash~\citep{Fishetal08}, except
that we use a degenerate equation of state, the same as that used for our
simulations of \SNeIa, with $\rho = \unit{7.3\ee{7}}{\grampercc}$ and $T
= \unit{4.3\ee{9}}{\K}$.  We drive the turbulence at a scale $L/3$ with
an energy to achieve $v_{\rm rms} \approx 0.1 c_s$, where $c_s \approx
\unit{6\ee{8}}{\cms}$ is the sound speed.  The turbulence cascade for all
simulations converge to a single steady-state profile in phase space by
$t = 1.5 \tau_e$.

We want to correct for numerical dissipation, so
we construct an idealized specific kinetic energy cascade 
that follows the well-known
$-5/3$ power law for steady-state, homogeneous and isotropic turbulence,
\ie, $E(k)
= A k^{-5/3} \erg\usp\cm\usp\power{\gram}{-1}$, where $k$ is the norm of the
vector wavenumber and $A$ is a proportionality constant. This energy
function represents the turbulence we would expect if we had infinite
resolution (and no numerical dissipation).  Intermittency is also being
ignored and corrections should be minor.  By definition, integrating the
energy function over all wavenumbers yields the total kinetic energy per
unit mass.
Because our idealized energy function is only valid between the
integral scale (\lt) and $\eta_k$ and most of the energy is contained at
the integral scale, we choose to solve for $A$ by approximating the integral
over all wavenumbers to those between $2\pi/\lt$ and 
$2\pi/\eta_k$
\begin{equation}
\int_{2\pi/\lt}^{2\pi/\eta_k} E(k) dk = \frac{1}{2} v_{\rm rms}^2 {\rm ,}
\end{equation}
where $\lt = L/3$, the scale at which turbulence is driven.
For the astrophysical flows of interest, $\ReN \sim 10^{14}$ such that 
$\eta_k \ll \lt$, and therefore, we take the upper bound of the integral to
infinity.
Then, defining $k_{\rm drive} = 2\pi / \lt$, $A$ becomes
\begin{equation}\label{eq:A}
A = \frac{1}{3} v_{\rm rms}^2 k_{\rm drive}^{2/3} {\rm ,}
\end{equation}
where $v_{\rm rms}$ is given in \tabref{tab:calibration} for all simulations. 
One could also choose to solve for $A$ by choosing an ``anchor'' value
within the inertial range of the cascade, but because an inertial range
is present only for the higher 512- and 1024-cell simulations, this method
may not be consistent for all cases.

By analogy with $v_{\rm rms}$, we want $\turbD$ to represent the turbulent
energy on unresolved scales, $E_{\rm SGS}(\turbD)={\turbD}^2/2$.
So instead of integrating from the driving scale, we 
integrate from the wavenumber corresponding to the length scale associated
with the turbulence operator, $k_\Delta = 2\pi/\Delta$
\begin{equation}\label{eq:Esgs}
E_{\rm SGS}(v_{\rm rms},\Delta) = \int_{k_\Delta}^\infty E(k) dk
=\frac{1}{2}v_{\rm rms}^2
\left(\frac{k_{\rm drive}}{k_{\Delta}}\right)^{2/3}
\end{equation}
The calibration constant is then derived from the desire for
$\langle E_{\rm SGS}(\turbD)\rangle = E_{\rm SGS}(v_{\rm rms},\Delta)$,
\begin{equation}\label{eq:c2}
\left(c_2\right)^2 = \frac{2 E_{\rm SGS}}%
{\left< {OP_2\left(\bvec{u}\right)}^2 \right>} =
\frac{v_{\rm rms}^2}{\left< {OP_2\left(\bvec{u}\right)}^2 \right>}
\left( \frac{k_{\rm drive}}{k_\Delta} \right)^{2/3}
{\rm .}
\end{equation}
While $\langle...\rangle$ is most appropriately an ensemble average
\citep{Davidson2004}, in the analysis below we use spatial averages over all
points in an assumed homogeneous box.
Here, $OP_2$ is log-normally distributed, as expected, with the standard
deviation being $\approx 5\%$ of the mean in log-space.

\begin{figure}[tb]
\centering
\includegraphics[width=\columnwidth]{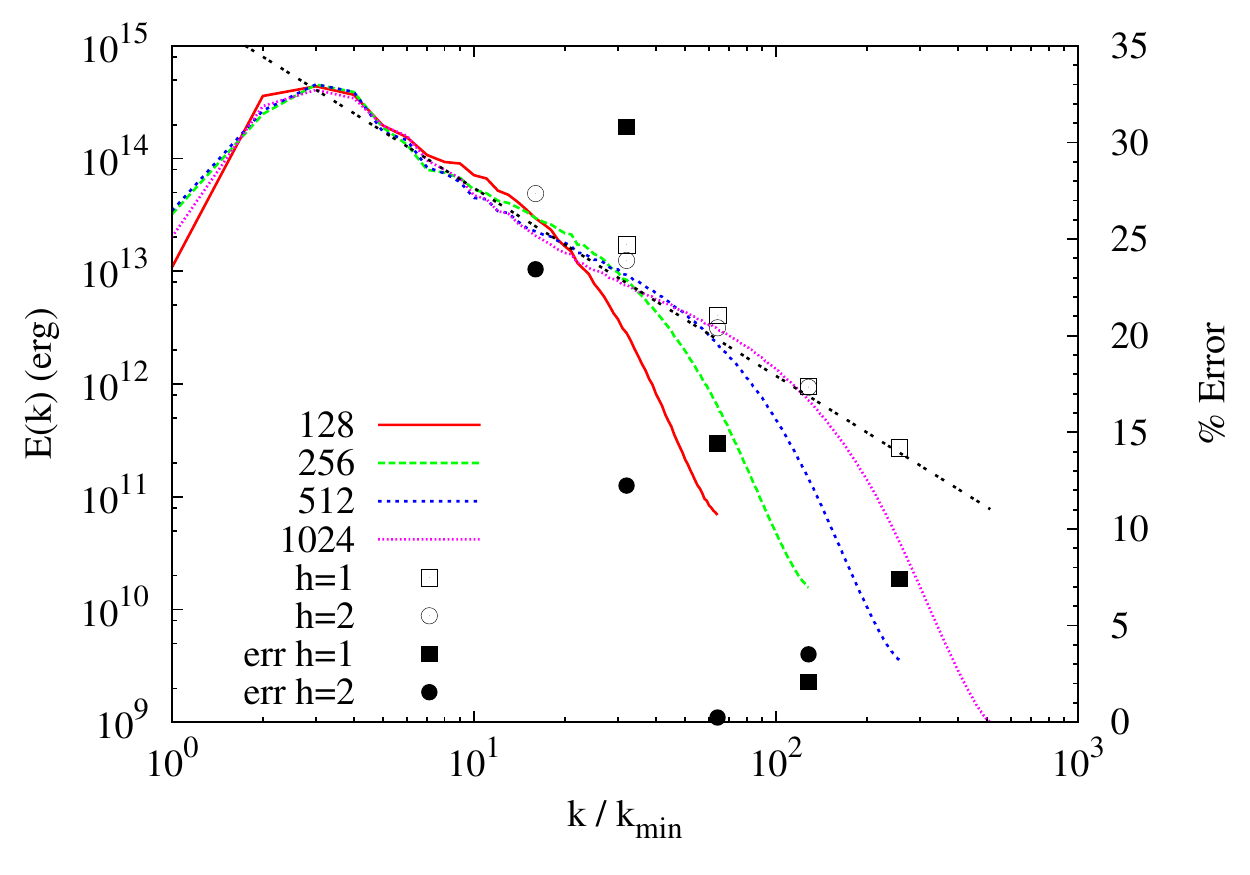}
\caption{The spectral energy content of stirred turbulence in a cubic
periodic box of size $L$ for four
different resolutions that range 128--1024 computational cells along each
dimension are shown as a function of wavenumber in units of $k_{\rm min}
= 2 \pi / L$. The 
open points show $E\left(k=k_\Delta\right)$, 
the energy contained in the length scale $\Delta$ (for different integer 
strides $h$), as inferred from the turbulence measurement operator (see text for details).
The dashed line is computed from $v_{\rm rms}$ with $E(k) = A k^{-5/3}$
and \eqref{eq:A} showing the expected $-5/3$ power-law. Because
each resolution results in a slightly different calibration for $c^h_2$,
we choose one representative value (provided in \tabref{tab:calibration})
and show the error in the turbulent
intensity with filled points associated with the right y-axis using
\eqref{eq:err}.
\label{fig:calibration}}
\end{figure}

\figref{fig:calibration} shows the
spectral energy content for four different resolutions, $N=$128, 256, 512
and 1024 cells along each direction.
The energy spectra, $E(k)$, are computed
by taking the Fourier transform of the velocity field and binning energy
in spherical shells of wavenumber $k = \sqrt{k_x^2 + k_y^2 + k_z^2}$. The
Fourier transforms are computed using the publicly available \texttt{FFTW}
routines,
with an additional multiplicative normalization factor of $1/N^3$.
From the energy spectrum, the
driving scale can be discerned as $k_{\rm drive} = 3 k_{\rm min}$, as
specified for the turbulence simulation.
The dashed line is computed from $E(k) = A k^{-5/3}$
and \eqref{eq:A} with $v_{\rm rms}$ from the 1024-resolution
simulation given in \tabref{tab:calibration}.
The large dissipation range can be observed, with a
well-defined inertial range only evident at $N=512$ and 1024.  These also
display the expected ``bump'' in the power spectrum between the inertial and
dissipation ranges~\citep{Sytietal00}.

Each resolution results in a slightly different value for $c^h_2$, and
these are presented in \tabref{tab:calibration}.  This is expected because
the nature of the dissipation range is known to depend on the overall
resolution \citep{Fishetal08,Aspdetal09}.  We choose a representative
value for each $h$, given in the last column of \tabref{tab:calibration},
and use this as our calibrated $c^h_2$.  Using this calibrated constant,
the open points in \figref{fig:calibration} show what the turbulent
operator infers the energy to be at $k_\Delta$, where $E(k_{\Delta}) = 2
E_{\rm SGS}(\turbD) / 3 k_{\Delta}$.
We show the error in the inferred turbulent intensity, incurred by using a
single, resolution-independent value for $c^h_2$, with filled points
associated with the right y-axis, where the error is given by
\begin{equation}\label{eq:err}
\textrm{\% err}\left( \turbD \right) = 
\frac{{\rm abs}\left( \turbD - \sqrt{2E_{\rm SGS}(v_{\rm rms},\Delta)} \right)}%
{\sqrt{2E_{\rm SGS}(v_{\rm rms},\Delta)}} \times 100 {\rm ,}
\end{equation}
which reduces to
\begin{equation}\label{eq:err2}
\textrm{\% err}\left( \turbD \right) = 
{\rm abs}\left( \frac{c^h_2}{c^h_2(N)} - 1 \right) \times 100 {\rm .}
\end{equation}
The error is found to be modest (less than 15\%) for a cascade that is
moderately well-captured on the grid, i.e. for resolutions of 256 and
above.  Even at lower resolutions, errors of 30\% are likely to be fairly
benign assuming that an important portion of the TFI is contained in the
resolved scales. Uncertainties of similar magnitude exist in the calibration
method itself. By choosing to ``anchor'' $A$ with an energy value within
the inertial range, $c^h_2$ for all cases drops by $\approx 0.1$,
resulting in a $\sim 10\%$ decrease for $h=1$ and $\sim 40\%$
decrease for $h=2$.

The calibration of $c^h_2$ presented in this work is different from
\citetalias{Colietal00} for a couple of reasons. First, \citetalias{Colietal00}
absorb the $h$ factor into $c^h_2$ whereas we include it explicitly in $OP_2$
in order to recover the appropriate scaling with length $\Delta$. Additionally,
\citetalias{Colietal00} use $h=2$ for a second-order finite central difference
approximation to the Laplacian, while using $h=1$ for the curl.

With the operator presented and calibrated here, we can form an
instantaneous model for the turbulence on unresolved scales.  This model
consists of a Kolmogorov cascade in which the overall magnitude of flow
velocities as a function of scale provides a smooth continuation of the
cascade from the resolved inertial range.  Though the velocity structure
is measured in the dissipation range, this has been compensated for in
order to provide a continuation of the inertial range.  With an additional
model for how a flame wrinkles in the presence of a Kolmogorov-like
velocity spectrum, we will be able to derive the behavior of the flame on
subgrid scales. This topic is pursued in the following sections.

\begin{table}[tbh]
\centering
\caption{Measured Values for $c^h_2$ for Various Resolutions $N$
with $v_{\rm rms}$
\label{tab:calibration}}
\begin{tabular*}{0.90\columnwidth}{@{\extracolsep{\fill}} l *{5}{r} }
\hline
\hline
N & 128 & 256 & 512 & 1024 & $c^h_2$ \\
\hline
$c_2(h = 1)$ & 0.69 & 0.79 & 0.92 & 0.97 & 0.9 \\
$c_2(h = 2)$ & 0.24 & 0.27 & 0.30 & 0.31 & 0.3 \\
\hline
$v_{\rm rms}$ (\kms) & 685 & 680 & 685 & 687 & \\
\hline
\end{tabular*}
\end{table}

\section{Turbulence--Flame Interaction Model}
\label{sec:tfimodel}

A turbulence--flame interaction (TFI) model estimates the turbulent
flame speed from the characteristics of the turbulent cascade and the 
laminar flame. For the TFI models discussed in this work, we limit our
analysis to combustion regimes in which the flame is a well-defined concept.
Generally for application to LES, TFI is concerned with accounting for
subgrid flame structure that enhances flame propagation above what the
resolved, and therefore necessarily smooth, reaction front achieves.
Models therefore predict the front propagation speed of the reaction front
resolved on scale $\Delta$, or the effective turbulent flame speed of the coarsened
reaction front, $s_{t\Delta}$.

\subsection{Expectations and Model Construction}
\label{sec:construct}

We implemented a few models based on \citetalias{Colietal00} and
\citetalias{Charetal02a}, finally deciding that \citetalias{Charetal02a}
is most complete and appropriate, out of those considered here, for
simulations of \SNeIa with \flash.
After some preliminary studies concerning how the models are constructed, our
adaptation of the power-law flame wrinkling model of
\citetalias{Charetal02a} is given in \secref{sec:powerlaw}.  After this,
as a supplement, we briefly discuss our attempts to directly adapt
\citetalias{Colietal00} which result in unsatisfactory models.
While we summarize the approaches here, see \citetalias{Colietal00} and
especially \citetalias{Charetal02a} for a full description.

\begin{figure}[tb]
\centering
\includegraphics[width=\columnwidth]{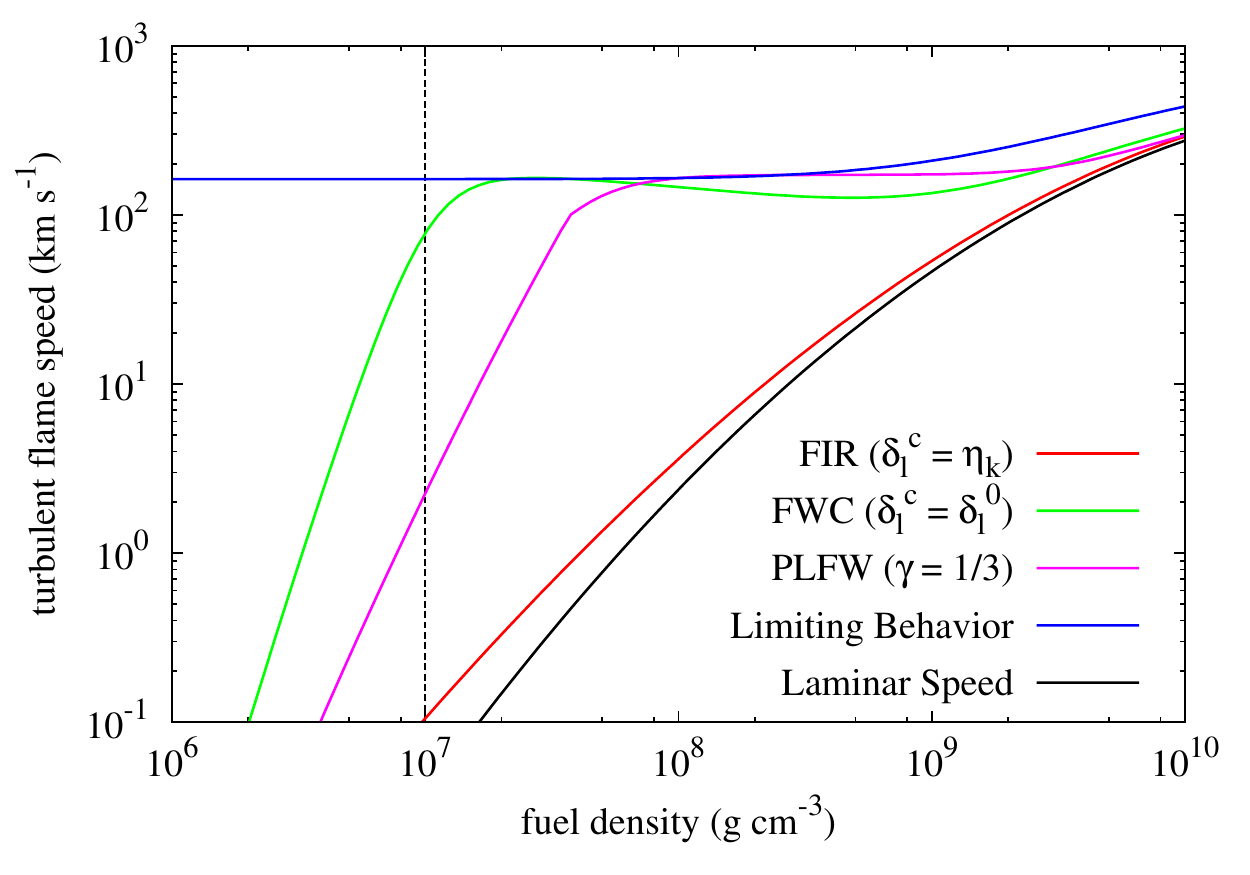}
\caption{
The turbulent flame speed is estimated using a simple Kolmogorov cascade
constructed with $\turb = \unit{300}{\kms}$ and $\lt = \unit{100}{\km}$,
where both
the laminar flame speeds and widths are log--log fits as functions
of density for a 50/50 C-O fuel mixture from \citet{ChamBrowTimm07}
with $\PrN = \power{10}{-5}$. The red and green lines are constructed
using the Full Inertial Range (FIR) and GS models with 
Equations~\ref{eq:alphaK} and
\ref{eq:alphaD}. The blue line shows the limiting behavior with 
$\Xi_\Delta = 1 + \turbD / \slamO$, while the black line is
the fitted laminar flame speed as a function of fuel density.
For comparison, results from the power-law flame wrinkling model 
using a $\gamma=1/3$
power-law are shown in magenta. In all cases, $\Delta = \unit{16}{\km}$.
The cut-off to $\sturbD$ that has typically been used in previous models
occurs at the suspected transition away from the
flamelet regime (vertical dashed line)~\citep{Schmetal06b,Townetal07}.
\label{fig:colin_st}}
\end{figure}

It is first useful to outline some expectations for our model from general
principles and known scaling laws.  Turbulent flames are often characterized
by a wrinkling factor $\Xi$, often defined as $\Xi \equiv A_t/A_l$, such that
$s_t=\Xi s_l^0$, where $s_t$ is the turbulent flame speed, $A_t$ is the surface
area of the turbulent flame, and $A_l$ is the surface area of the laminar flame.
For clarity, we are following the convention in CVDP and using $\slamO$ for
the laminar flame speed of the physical flame, for which we have simply used
$s_l$ in previous sections. This is intended to contrast with $s_l^1$, which
would be the speed of the thickened or artificial flame, which is our ADR
front.
The wrinkling factor describes the increase in flame surface area due to
turbulence.  We leave
the length scale on which this might apply slightly abstract for the moment.
One well-known scaling law is Damk\"ohler scaling, in which, for
moderately strong turbulence, the average flame speed 
$s_t\approx \turb$, where $\turb$ is the characteristic turbulence intensity.
One way to parameterize this scaling is to suggest a
wrinkling factor of the form
\begin{equation}
\label{eq:damkohler}
\Xi = 1+\beta\frac{\turb}{s_l^0}\ ,
\end{equation}
where $\beta$ is a constant of order 1.  For the \SNIa, the physical
laminar flame speed,
$s_l^0$ is strongly dependent upon fuel density.  It is therefore
convenient to consider the dependence shown in \figref{fig:colin_st},
where $s_t$ is plotted against density for a given turbulence field.  In
keeping with the rest of the discussion, we emphasize $s_{t\Delta}$,
the enhancement to subgrid propagation.
A simple Kolmogorov cascade is assumed with $\turb = \unit{300}{\kms}$ and
$\lt = \unit{100}{\km}$, where both the laminar flame speeds and
widths are log--log fits as functions of density for a 50/50 C-O 
fuel mixture from \citet{ChamBrowTimm07} with $\PrN = 10^{-5}$, where $\PrN$
is the Prandtl number that describes the ratio of kinematic viscosity
to thermal diffusivity.
The blue line shows the limiting behavior with $\Xi_\Delta = 1 + 
\turbD / \slamO$, while the black line is the fitted laminar flame speed.
These estimates were calculated with $\Delta = \unit{16}{\km}$, which is
approximately the length scale associated with the turbulence measure for
$h=1$ in a $\unit{4}{\km}$ resolution simulation.

This scaling, however, does not extend to infinitely low $s_l^0$.  At some
stage $\Xi$ is as high as it can be given the finite width of the flame and
the volume it represents,
and the Damk\"ohler relation is no longer followed.  As a result, as
$s_l^0$ decreases and $\delta_l^0$ increases, $s_{t\Delta}$ should fall
off at low densities.  The cut-off to $\sturbD$ that has typically been
used in previous models~\citep{Schmetal06b, Townetal07} is given by the
vertical dashed line in \figref{fig:colin_st}. This line corresponds to estimates of the density at
which combustion in a degenerate WD transitions from the flamelet regime
to the ``distributed burning'' or ``broken reaction zone'' regime. This
estimate is typically derived from the condition that $\gibson \lesssim
\dlamO$, where $\gibson$ is the Gibson scale that describes the length
scale at which the turbulent eddies are burned by the flame in less than an
eddy turnover time. If $\gibson$ drops below $\dlamO$, turbulent eddies
can turnover inside the flame structure and contribute to the transport
of mass and heat. If this transition occurs, the flame is no longer a
well-defined concept, and the assumptions used in the construction of 
the models for turbulent combustion are no longer valid.

The general concept explored by both \citetalias{Colietal00} and
\citetalias{Charetal02a} based on earlier work by \citet{MenePoin91} is to
integrate the wrinkling over many scales (through the cascade) in order to
consider separately the contributions from above and below the approximate
LES smoothing scale $\Delta$.  At the heart of this method is the
consideration of single-scale flame-vortex interactions.
Using direct numerical simulations (DNS) of single flame--vortex
interactions, \citetalias{Colietal00} investigate the efficiency of the
vortex to wrinkle the flame when the ratio of the vortex size ($r$) to
laminar flame width and ratio of the vortex velocity ($v^\prime$) to the
laminar flame speed varies. A subgrid scale model is constructed by
calculating the effective strain rate on the flame by integrating the
efficiency of single flame--vortex interactions (computed from DNS)
through the unresolved inertial range of the turbulent cascade.
This introduces two important questions: how does the contribution to
wrinkling depend on scale? and what is the appropriate inner cutoff scale
under various conditions?  These will be answered in different ways by the
models discussed.

\subsection{Power-Law Flame Wrinkling Model}
\label{sec:powerlaw}

\citetalias{Charetal02a} focus on modeling the wrinkling of the unresolved
physical flame. They consider a flame that is fractal in nature, with an
inner scale determined by the characteristics of the turbulence and flame.
They also consider the possibility that the fractal dimension could depend
on other local properties, but we do not include this part of their
analysis~\citep{Charetal02b}.
\citetalias{Charetal02a} postulate that the inner length scale is the
inverse mean curvature of the flame, which they solve for directly using
the flame surface density balance equation and the hypothesis that flame
surface destruction and production are in equilibrium on unresolved scales.
This construction requires only grid-scale quantities to be known, an
advantage over the methods discussed by \citetalias{Colietal00} (see
\secref{sec:colinetal}).  The only stipulation to recover Damk\"{o}hler
scaling is that, in this limit, the fractal dimension of the flame should
be $D = 7/3$.

\citetalias{Charetal02a} obtain their power-law flame wrinkling model
from the postulate that the wrinkling factor is a simple power function
involving the dimensionless ratio of the smoothing scale $\Delta$ to the
inner cutoff scale of the flame structure $\eta_c$.  This is expressed by the
wrinkling factor given in
\citetalias[\eq{3}]{Charetal02a}:
\begin{equation}\label{eq:CMV3}
\Xi_\Delta = \left( 1 + \frac{\Delta}{\eta_c} \right)^\gamma {\rm,}
\end{equation}
where then $s_{t\Delta}=\Xi_\Delta s_l^0$, and
$\gamma$ is not necessarily constant. If $\gamma$ is independent
of scale and restricted to $0 < \gamma < 1$, the fractal model is recovered
where the fractal dimension $D = \gamma + 2$. In the limit of strong
turbulence, $\gamma = 1/3$ in order to recover Damk\"ohler scaling as shown
in \figref{fig:colin_st}. The inner cut-off scale
$\eta_c$ is estimated to be the inverse mean curvature of the flame
($\meancurve^{-1}$).
This is then obtained by assuming equilibrium between subgrid flame
surface creation (strain) due to wrinkling by the turbulence and flame
surface destruction by flame surface propagation and diffusion.
This gives \citetalias[\eq{10}]{Charetal02a}:
\begin{equation}\label{eq:CMV10}
\eta_c^{-1} =
\meancurve = \Delta^{-1} \us \Gamma\left(\Dd,\us,\ReD\right) {\rm ,}
\end{equation}
where $\Gamma(..)$ is an efficiency function which takes into account the
net straining effect of all relevant turbulent scales smaller than
$\Delta$, and is derived by integrating over the single-vortex efficiency
function, $C$ \citepalias{Colietal00}, and a turbulence spectrum (see
below).
The inverse mean curvature is related to the Gibson scale, although it is
not immediately obvious.  See the discussion of the scaling in various
regimes in \citetalias{Charetal02a}.

\citetalias{Charetal02a} prevent the unphysical result, $\eta_c < \dlamO$,
with
\begin{equation}\label{eq:inner-cut}
\eta_c = \max{\left(\meancurve^{-1},\dlamO\right)} {\rm .}
\end{equation}
This limiting behavior provides a natural and physical way to quench
flames. Even for a space-filling flame with $\gamma = 1$, there is a
finite amount of flame surface that can physically exist on unresolved
scales for a finite-width flame.
Models currently in use, including
\citet{Khok95} and \citetalias{Schmetal06b}, do not consider this effect,
which may be important for combustion in \SNeIa\ for the density range
$\unit{\power{10}{7}-\power{10}{8}}{\grampercc}$. Instead, these
approaches typically set $\sturbD = 0$ for 
$\rho < \unit{10^7}{\grampercc}$, where $\rho$ is the fuel density.

\begin{figure}[tb]
\centering
\includegraphics[width=\columnwidth]{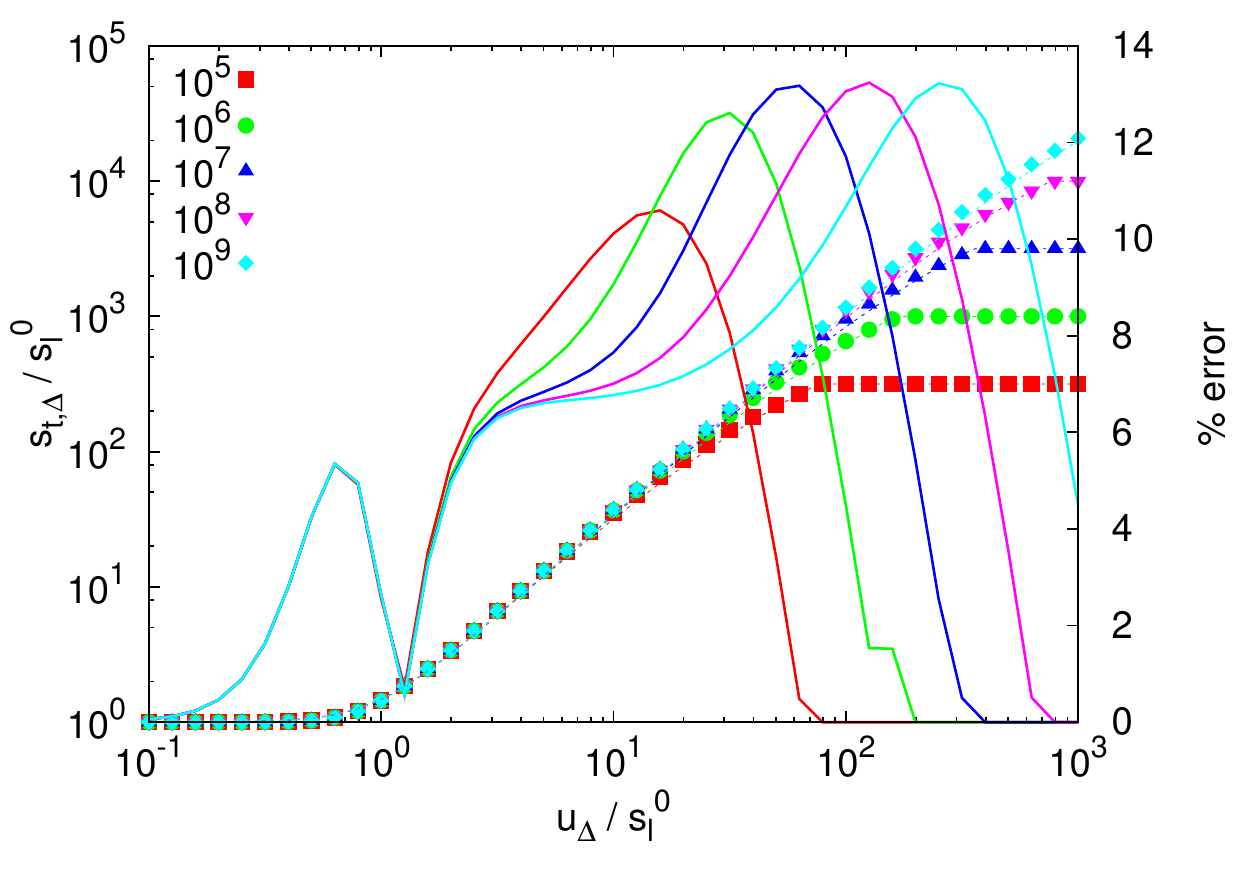}
\caption{
The enhancement to the laminar flame speed for $\PrN = 10^{-5}$
is shown as a function of the ratio of the turbulent intensity at the
scale $\Delta$ to the laminar flame speed ($\turbD/\slamO$) on the x-axis
and the ratio of $\Delta$ to the laminar flame width ($\Delta/\dlamO$)
with different colors. The numerically integrated values are shown as
points while the fit values are shown as thin dashed lines. The percent
error between the fit and the numerically integrated result are shown as
thick solid lines with the error given on the second y-axis. The range
of values chosen represents the range of expected values for a moderate
resolution \SNeIa simulation. The fit-errors are all $\lesssim 10\%$.
\label{fig:lowPr_enhance}}
\end{figure}

The spectral efficiency function $\Gamma$ utilizes an
efficiency function $C$ modified from \citetalias{Colietal00},
such that vortices with speed
$v' < \slamO/2$ do not wrinkle the flame. 
Additionally, the
integration over length scales is performed in $k$-space in
\citetalias[\eq{13}]{Charetal02a}:
\begin{equation}\label{eq:CMV13}
\left(\Gamma \frac{\turbD}{\Delta} \right)^2 = \int_{\pi/\Delta}^\infty 
\left[C\left(k\right)\right]^2 k^2 E_{11}\left(k\right) dk
{\rm ,}
\end{equation}
where $E_{11}(k)$ is the one-dimensional energy spectrum describing
homogeneous, isotropic turbulence in the 
direction of the wavenumber $k$, given in \citetalias[\eq{16}]{Charetal02a}. 
Because $E_{11}(k)$ depends on
\PrN, we re-evaluate the numerical integral for
$\PrN = 10^{-5}$ to compare with fitting functions provided
by \citetalias{Charetal02a} in \figref{fig:lowPr_enhance}. Numerical
integrals are computed using \citetalias[\eqs{18--21}]{Charetal02a} and
\eqref{eq:Re} with subroutine \texttt{dqag} from the
publicly available \texttt{SLATEC} library. Because the integrand is
non-oscillatory, smooth and tends toward zero exponentially as the
argument goes to infinity, we choose an upper bound to the integral
such that the integrand evaluates to a number less than $10^{-10}$.
This criterion is sufficiently small as compared to the integrand
evaluated at the lower bound. A good initial guess at the upper bound
is given by $10^{3}$ multiplied by \citetalias[\eq{26}]{Charetal02a}, which
approximates the maximum of the integrand. 
We evaluate $\Gamma$ with
an error tolerance of $0.1\%$ and found that a Gauss--Kronrod pair with 
$15-31$ points provides consistent results for a wide range of parameter
space. We also use the fitting function for $\Gamma$ given in 
\citetalias[\eqs{30--34}]{Charetal02a} in a range outside that
explored by \citetalias{Charetal02a}. We compare
numerically integrated values (points) to the
fit (thin dashed lines) for $\Delta/\dlamO = \power{10}{5}-\power{10}{9}$ and a 
range of
$\turbD/\slamO$ up to 
$\power{10}{3}$. The associated percent errors are provided as thick lines
in \figref{fig:lowPr_enhance}.

For comparison, results from the power-law flame wrinkling model 
with $\gamma=1/3$ corresponding to the 
limiting Damk\"ohler scaling behavior are shown in magenta in
\figref{fig:colin_st}.
Given the arguments by \citetalias{Charetal02a} that $\Xi^1 \approx 1$ and
the ambiguities associated with choosing a cut-off scale $\delta_l^c$ in
the models derived below from \citetalias{Colietal00}, we choose
to implement the power-law flame wrinkling model. The power-law flame 
wrinkling model provides a more natural mechanism
to quench turbulent flames rather than supplying an ad-hoc prescription
as in other models~\citep{Schmetal06a,Townetal07}. Additionally, the
implementation requires only grid-scale quantities, which are computationally
accessible.

\subsection{Other Models Considered: a priori cutoffs}
\label{sec:colinetal}

\citetalias{Colietal00}
specifically investigate a model applicable to artificially thickened
flames. In order to achieve a thickened flame resolvable by the
computational domain, the thermal and molecular diffusivities are
enhanced by a thickening factor $F$, while the reaction rate is reduced
by $F$. However, by increasing the diffusivity, the thickened flame is
less responsive to strain on the flame front due to turbulence. Thus, 
while the flame is resolved on the domain, the interaction with
turbulence is affected. Because the utilization of artificially thickened
flames greatly reduces computational costs while capturing many of
the inherent phenomena associated with combustion,
\citetalias{Colietal00} attempt to account for inefficient wrinkling of
the artificially thickened flame by resolved turbulence.  Thus the
formalism tries to consider the context in which both the artificial and
real flames exist, and correct the former to the latter.
Since it is posed as a relative correction, rather than an absolute
calculation of surface area, the wrinkling factors must be normalized to
reproduce some limiting behavior, taken to be Damk\"{o}hler scaling
($\sturb \propto \turb$).  Posed in this way, the resulting normalization
requires integral scale quantities to be known, such as
the integral scale ($\lt$), turbulent intensity ($\turb$), and Reynolds
number (\ReN).  This also creates the situation where an a priori choice
must be made for the appropriate small-scale cut-off under which
Damk\"{o}hler scaling is recovered.

In \citetalias{Colietal00}, the enhancement to the laminar flame speed is given
by the ratio of the wrinkling factor for the thin flame to the wrinkling
factor for the thick flame, $E = \Xi^0 / \Xi^1$. Here, the superscript $0$
refers to the real physical ``thin'' flame, and the superscript $1$ refers 
to the model ``thick'' flame.
Here now $s_{t\Delta}=Es_l^0$, since $\Xi^1\ne1$.
The wrinkling factor is given by equating
subgrid flame surface production from turbulent wrinkling to subgrid flame
surface destruction by propagation and diffusion from the subgrid
flame surface density balance equation in \citetalias[\eq{18}]{Colietal00}:
\begin{equation}\label{eq:CVDP18}
\Xi \approx 1 + \alpha \frac{\Delta}{\slamO}\left<a_T\right>_s
{\rm ,}
\end{equation}
where $\alpha$ is a model constant, $\Delta$ is the length scale associated
with the turbulence operator, and $\left<a_T\right>_s$ is the effective
strain rate averaged over the flame surface on subgrid scales. The 
effective strain rate is given by \citetalias[\eq{24}]{Colietal00}
calculated by integrating turbulent motions over the inertial range by an
efficiency function derived from DNS calculations:
\begin{eqnarray}
\label{eq:CVDP24}
\left<a_T\right>_s &=& \frac{c_{ms}}{\ln{\left(2\right)}}
\int_{\rm scales} C \left( \frac{r}{\delta_l}, \frac{v^\prime}{\slamO} \right)
\frac{v^\prime}{r} d \left[ \ln{\left(\frac{\lt}{r}\right)} \right] {\rm ,}
\\
&=& \Gamma\left(\frac{\Delta}{\delta_l},\frac{\turbD}{s_l^0}\right)
\frac{\turbD}{\Delta}
\end{eqnarray}
where $c_{ms}$ is a constant determined from DNS 
calculations~\citep{Yeunetal90} and $\slamO$ refers to the unstrained 
laminar flame speed (superscript $0$ referring to unstrained).
Note that the efficiency function and the integration differ slightly from
\citetalias{Charetal02a}.
The result, $\Gamma$, of performing this integration with the DNS-derived
efficiency function, $C$, is given by \citetalias[\eq{30}]{Colietal00}.
This leaves $\alpha$ to be determined.
In order
to evaluate the
limiting case of Damk\"{o}hler scaling ($\Xi \approx 1 + \beta
\turb/\slamO$), a simple Heaviside efficiency
function is introduced for $C$ with a cut-off scale $\delta_l^c$, such that
$C=0$ for $r < \delta_l^c$ and $C=1$ for $r > \delta_l^c$. In this
case, the integral can be solved analytically with the caveat that
$\delta_l^c \ge \eta_k$:
\begin{equation}\label{eq:alpha}
\alpha = \beta \frac{2 \ln{\left(2\right)}}{3 c_{ms}}
\left[ \left(\frac{\lt}{\delta_l^c}\right)^{2/3} - 1 \right]^{-1} {\rm ,}
\end{equation}
where $\beta$ is a model constant of order unity. We describe two
procedures to calculate $\alpha$ that lead to two different models.

\subsubsection{Full Inertial Range Model}
\label{sec:fullmodel}

\citetalias{Colietal00} argue Damk\"{o}hler scaling is achieved only
when the flame front is wrinkled by \emph{all} turbulent motions
in the inertial range, from $\eta_k$ to $\lt$. This hypothesis defines
the model we call the ``Full Inertial Range'' (FIR) model. For astrophysical
flames, typically $\eta_k \ll \dlamO$, which means that $\alpha$ will be relatively small, and the model will not predict much enhancement.

\citetalias{Colietal00}
let $\delta_l^c = \eta_k$ to evaluate $\alpha$ in \eqref{eq:alpha}.
When $\delta_l^c = \eta_k$, $\left(\lt/\delta_l^c\right)^{2/3} = 
\ReN^{1/2}$. Here, the evaluation of \ReN becomes important, where
$\ReN = \lt \turb / \nu$, and integral scale quantities are necessary.
We prefer to write \ReD for scale $\Delta$
in terms of the ratios $\turbD/\slamO$ and $\Delta/\dlamO$
\begin{equation}\label{eq:Re}
\ReD = \us \Dd \frac{\slamO \dlamO}{\nu} = \us \Dd \PrN^{-1} {\rm ,}
\end{equation}
where $\slamO \approx \sqrt{\kappa / \tau_r}$ and $\dlamO \approx 
\sqrt{\kappa \tau_r }$, such that $\slamO \dlamO = \kappa$, where 
$\kappa$ is the thermal diffusion coefficient and $\tau_r$ is the 
reaction timescale. Then, $\PrN = \nu / \kappa$. Here, $\dlamO$ 
refers to the physical laminar flame width; whereas, $\dlam^1$ will
refer to the model flame width. \ReN 
at the integral scale can be evaluated with $\Delta = \lt$. For 
terrestrial flames, $\PrN \sim 1$ with \citetalias{Colietal00} choosing 
$\PrN = 1/4$; however, astrophysical flames characteristic for a
degenerate WD have \PrN as low as $10^{-5}$~\citep{NandPeth84,
NiemKers97,Kers01}.
In our simulations, integral scale quantities are not easily accessible
and it is useful to solve for $\alpha$ at the scale $\Delta$ such that
\begin{equation}\label{eq:Xi}
\Xi_\Delta = 1 + \alpha \frac{\Delta}{\slamO} \left<a_T\right>_s = 
1 + \beta \us {\rm .}
\end{equation}
This means that the turbulent flame speed at the scale $\Delta$ obeys
Damk\"ohler scaling when all turbulent motions
of size $r$ with $\delta_l^c < r < \Delta$ wrinkle the flame with full efficiency.
For the implementation in which $\delta_l^c = \eta_k$, we obtain
\begin{equation}\label{eq:alphaK}
\alpha_{\rm FIR} = \beta \frac{2 \ln{\left(2\right)}}{3 c_{ms}}
\ReD^{-1/2} {\rm .}
\end{equation}
For low-\PrN flames, we never actually
recover $\Xi_\Delta = 1 + \beta \turbD/\slamO$ because $\eta_k < \dlamO$.
As seen from the red curve in \figref{fig:colin_st}, $s_{t\Delta}$ is never
much more than $s_l^0$, which does not meet our expectations for the
behavior of the turbulent flame speed in these circumstances.
For these reasons, we consider another normalization to solve for $\alpha$.

\subsubsection{Flame Width Cutoff}
\label{sec:GS}

Instead of only allowing Damk\"ohler scaling when the flame is wrinkled
by the full inertial range, we allow Damk\"ohler scaling when the flame
is influenced by all turbulent eddies ranging from $\Delta$ to the flame
scale, letting $\delta_l^c = \dlamO$. This hypothesis defines the model
we call the "Flame Width Cutoff" model.
With $\Delta/\dlamO \gg 1$, the evaluation of $\alpha$ becomes
\begin{equation}\label{eq:alphaD}
\alpha_{\rm{Flame Width Cutoff}} = 
\beta \frac{2 \ln{\left(2\right)}}{3 c_{ms}} \left( \Dd \right)^{-2/3}
{\rm .}
\end{equation}

The resulting $s_{t\Delta}$ is shown by the green line in
\figref{fig:colin_st}, and provides a reasonable approximation of the
expected behavior.  The apparent physical underpinnings for this are,
however somewhat suspect.  We have made the assumption that Damk\"ohler
scaling holds whenever the turbulence cascade reaches the flame width.
This creates the unphysical result that when $\delta_l^0$ starts to become
large, the model assumes that the resolved flame will capture the
Damk\"ohler scaling, and the enhancement falls off.  This is therefore not
a physically consistent model, despite having the expected behavior, and
leads us again to prefer the construction of \citetalias{Charetal02a}.

\section{Verification}
\label{sec:verification}

We present two formal verification \citep{calder_2002_aa} tests of our TFI
model as a whole.  The whole model includes the turbulence measurement
operator described in \secref{sec:turbulence}, the TFI prescriptions
developed from the power-law flame wrinkling model of \citetalias{Charetal02a}, as presented in
\secref{sec:powerlaw}, and a ADR scheme for propagating the reaction
front, described in previous work \citep{VladWeirRyzh06,Townetal07}.
We wish to
verify by numerical tests that the TFI model produces sensible results in
two regimes: negligible turbulence and moderate-strength freely decaying
homogeneous turbulence.  These are the most important regimes for the
early part of the \SNIa.  We constrain ourselves to the type of turbulence
considered in construction of the model, leaving more complex turbulence
fields to future work.

\subsection{TFI Channel Test}

It is important to develop flow configurations that are simultaneously
numerically tractable and provide meaningful verification of TFI
techniques.  Those performed in the astrophysical literature so far have
used flames in a fixed-size box of stirred turbulence (\citetalias{Schmetal06a}, see also
\citealt{Schmthesis04}), a flame propagating up a vertical channel with
self-created turbulence due to the RT instability
\citep{Khok93,Khok95,Zhanetal07,Townetal08}, and a flame propagating along
a channel into a driven turbulent velocity field \citet{Aspdetal08b}.
Motivated by \citetalias{Charetal02a}, here we develop and utilize a
configuration based on a flame propagating along a channel into a field of
decaying turbulence.  Our basic configuration is shown in
\figref{fig:overview}, in which a flame is allowed to propagate
longitudinally along a (laterally periodic) channel against the prevailing
flow.  A turbulent field, obtained from a stirred turbulence simulation,
is placed immediately ahead of the flame and allowed to decay as the flame
propagates into it.

In order to test our implementation of the power-law flame wrinkling
subgrid scale TFI, we use a simple 1-stage model flame described in
\citet{VladWeirRyzh06,Townetal07}. The model flame is
described by a reaction progress variable, $\phi$, that is evolved by an
advection-reaction-diffusion equation where the reaction and diffusion
terms are chosen to yield a specified front propagation speed and model
flame width~\citep{Khok95}. The use of a 1-stage burner with a known
energy release results in an analytic solution to the Rankine-Hugoniot
jump conditions across the flame used to set up the initial conditions.
For our tests, our flame burns a 50/50 carbon-oxygen mixture by mass
to 50/50 oxygen-magnesium, which is equivalent to the simplified prescription
of the carbon burning stage in our 3-stage flame model that is designed to 
capture the energetics in \SNeIa~\citep{Townetal07}. For
simplicity, we specify constant laminar flame speed properties with
a speed of $\unit{10}{\kms}$ and physical width of 
$\unit{\power{10}{-3}}{\cm}$.

The flame is
initialized in the center of a box that is four times longer than it is wide
using Cartesian coordinates with the model flame resolved using four zones
propagating in the positive direction of the first dimension. Subsonic
inflow and outflow boundary conditions motivated by \citet{PoinVeyn05}
are imposed. A subsonic inflow boundary condition is used with an inflow
velocity initially equal to the specified front propagation speed such
that the model flame remains in the center of the box. The inflow velocity
is allowed to vary for turbulent flows such that fuel is in-flowing at
the burning rate
\begin{equation}\label{eq:inflow}
v_{\rm inflow} = \frac{\dot{m}_b}{A_{\rm inflow} \rho_{\rm fuel}} {\rm ,}
\end{equation}
where $v_{\rm inflow}$ is the inflow velocity defined to be positive in
the $-x$ direction, $\dot{m}_b$ is the burning rate and $A_{\rm inflow}$
is the cross-sectional area of the inlet. The burning rate is calculated
from the change in mass of ash on the grid between time steps and the
ash outflow rate
\begin{equation}\label{eq:brate}
\dot{m}_b = \frac{m_{\rm ash}^n - m_{\rm ash}^{n-1}}{t^n - t^{n-1}} +
   \left< \rho v_{\rm outflow} \right> A_{\rm outflow} {\rm ,}
\end{equation}
where $v_{\rm outflow}$ is the outflow velocity defined to be positive
in the $-x$ direction, $A_{\rm outflow}$ is the cross-sectional area
of the outlet, $m_{\rm ash} = \int\phi\rho\,dV$ and $t^n$ denotes
the $n$-th time step.  A static-pressure subsonic outflow boundary
condition is imposed that conserves mass and energy through the
boundary using the steady-state Euler equations and divergence theorem
assuming zero-gradient velocities in orthogonal directions. Because the
steady-state continuity equation is satisfied, $\rho v_{\rm outflow}$
is constant across the boundary allowing an average, $\langle \rho v_{\rm
outflow}\rangle$, over the interior
boundary cells.  A static-pressure outflow reflects sound waves back
into the computational domain, which are necessary to stabilize the
flow.  The inflow temperature is strictly maintained, while the inflow
velocities are allowed to vary to ensure acoustic reflections back into
the computational domain are dampened. If perfectly reflecting boundary
conditions are used on both the outflow and inflow, sound waves never
leave the computational domain.  Periodic boundary conditions are imposed
in the second and third dimensions.

\begin{figure}[tb]
\centering
\includegraphics[width=\columnwidth]{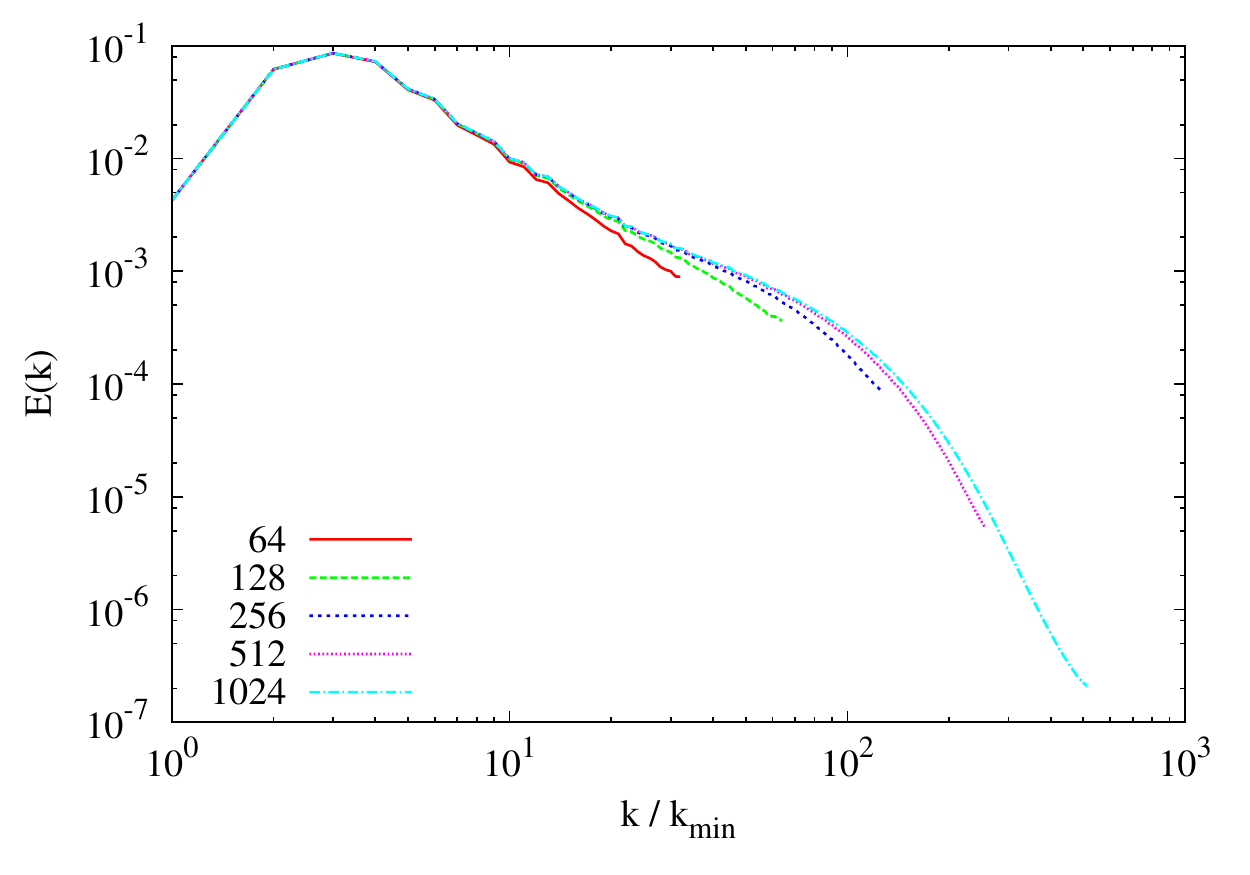}
\caption{
The normalized spectral energy content for smoothed velocity fields resulting
from the $\power{1024}{3}$-resolution turbulence run in
\figref{fig:calibration} as
a function of wavenumber in units of $2\pi/L$.
\label{fig:smoothed}}
\end{figure}

Our verification tests include simulating a laminar flame in a
channel with varying resolutions from 64x64 to 256x256 zones in
cross-sectional resolution.
Cells are cubic and the total simulated domain is twice as long in the
longitudinal direction as it is in the transverse direction.
While these resolutions are possibly too low to capture a true inertial
range, they are similar to conditions in the supernova that are challenging
for the same reason.
Because the initial profile given for the
reaction progress variable, $\phi$, is not equal to the true steady
state profile, it takes a few flame self-crossing times, $\tau_{\rm flame}
= \dlam^1/\slamO$, to obtain a steady
state solution. Firstly, we verify that for perfectly laminar flow,
the TFI model recovers the laminar flame speed as expected. Afterwards,
we restart the simulation by superimposing a turbulence field with a
specified turbulent intensity in the fuel. The turbulence field used is
the same as the $\power{1024}{3}$-resolution run used to calibrate the constant
$c_2$ in \secref{sec:turbulence} with the velocities normalized
by $v_{\rm rms}$. In addition, velocities are smoothed via direct
averaging to obtain lower-resolution turbulent velocity fields. This
procedure allows a direct comparison of the TFI model as the inertial
range of the turbulent cascade is better resolved. The power spectra of
the smoothed velocity fields is provided in \figref{fig:smoothed}.
These
simulations are computed for various resolutions to verify that the
combined effects of resolved and unresolved TFI result in a consistent
overall turbulent flame speed independent of the grid resolution.

In each of these tests, the flame speed is computed with
\begin{equation}\label{eq:flamespeed}
s_A = \frac{\dot{m}_b}{A \left<\rho_{\rm fuel}\right>} {\rm ,}
\end{equation}
where $s_A$ is the flame speed computed from using area $A$, which is
the surface area of an iso-contour of the reaction progress variable
to compute the front-propagation speed or the cross-sectional area
of the channel to compute the turbulent flame speed. In the case of
an iso-contour, the surface area is computed using the marching cubes
algorithm.
We estimate
the fuel density by assuming a low-Mach number isobaric burn and solving
for $\rho\left(\phi=0\right)$ from~\citep{VladWeirRyzh06}
\begin{eqnarray}
e\left(p,\rho\right) + \frac{p}{\rho} + q\phi &=& \emph{const} \\
p &=& \emph{const} {\rm ,}
\end{eqnarray}
where $q$ is the total energy released through the burn, $p$ is the
pressure, and $e$ is the specific internal energy determined from the
equation of state. The averaging procedure to obtain $\langle\rho_{\rm
fuel}\rangle$ is performed on densities just ahead of the flame with $10^{-6}
< \phi < 5\ee{-2}$.

\subsection{Recovering the Laminar Flame Speed}
\label{sec:laminar}

\begin{figure}[tb]
\centering
\includegraphics[width=\columnwidth]{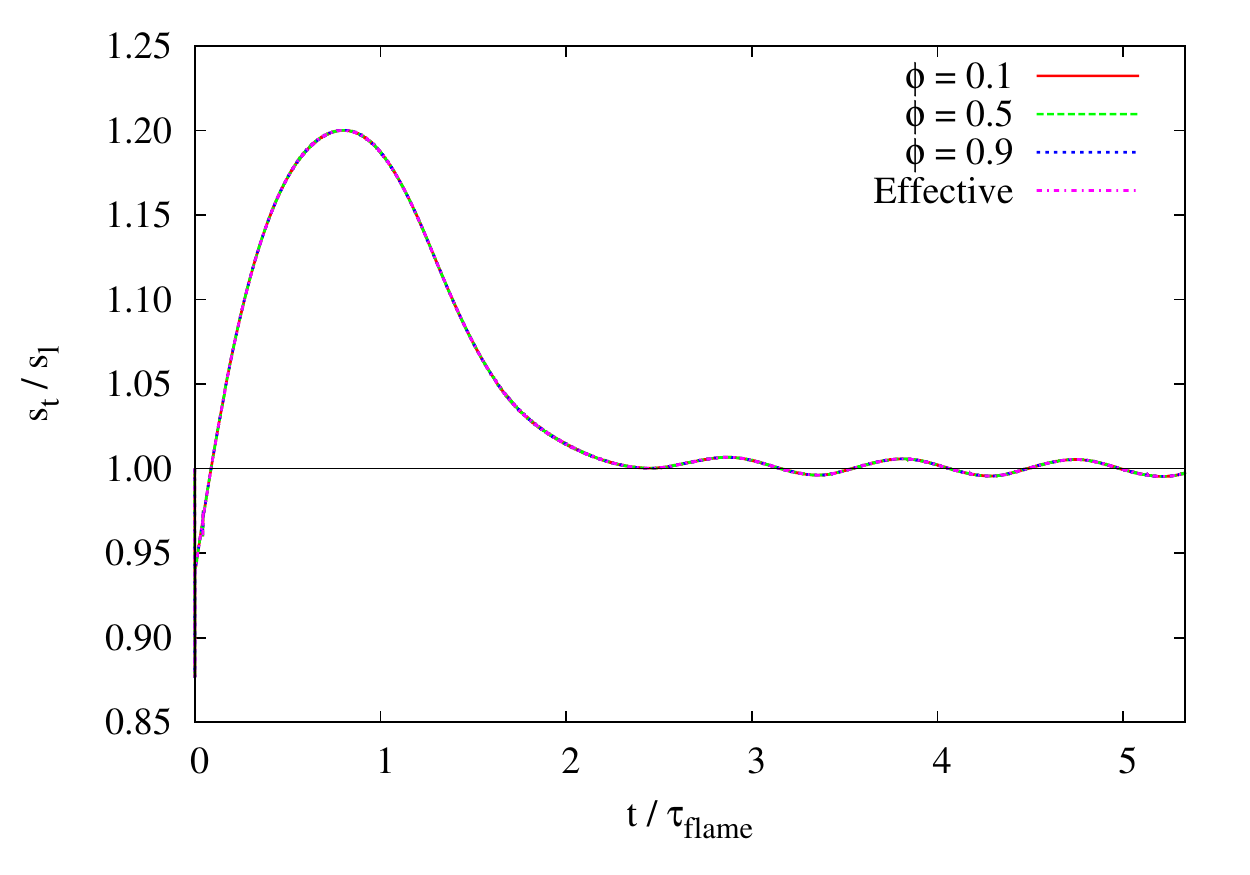}
\caption{
The ratio of the front propagation speed to input speed of the 
iso-contours of the reaction
progress variable for $\phi = \{0.1, 0.5, 0.9\}$ (red, green, and
blue lines, respectively) as well as the
turbulent flame speed (magenta) are compared. The initial flow is 
computed by solving the Rankine-Hugoniot
jump conditions in the reference frame of the flame such that
the flame remains in the center of the domain~\citep{VladWeirRyzh06}.
The small oscillations in the estimated flame speed are discussed in
\citet{Townetal07}.
\label{fig:laminar}}
\end{figure}

We test the implementation of $\turbD$ in \eqref{eq:turbD} and the 
power-law flame wrinkling subgrid scale TFI model in the limit
of $\turbD / \slamO \to 0$. In this limit, the turbulence operator
is constructed to ignore the expansion of material due to laminar
flame propagation and $\turbD = 0$ should be observed. In addition,
the TFI model should calculate the enhancement $\Xi_\Delta = 1$ such that
$\sturbD = \slamO$.  For this test, we fix the inflow velocity equal
to the input front-propagation speed and do not allow it to vary. In
\figref{fig:laminar}, we plot the ratios of the front-propagation speeds
to the input speed of the iso-contours of the reaction progress variable
for $\phi=\{0.1, 0.5, 0.9\}$ as well as the turbulent flame speed compared.
All estimated flame
speeds are perfectly consistent with one another indicating that the
area of each iso-contour and the cross-sectional area of the channel
are all equivalent. The initial adjustment and subsequent oscillation of 
the $\phi\left(x\right)$ profile is related to the flame self-crossing
time $\tau_{\rm flame} = \dlam^1 / s$, which is described in detail in
\citet{Townetal07}.  This shows that the turbulence
operator indeed measures $\turbD=0$ and that the resulting flame speed
is equivalent to the laminar (or input) flame speed.  The Landau-Darrieus
instability is not observed because the initial profiles are sufficiently
flat that it does not grow significantly in this short timeframe.

\subsection{Convergence Study}
\label{sec:convergence}

By increasing the resolution of the computational domain, more of the
turbulent cascade is resolved and wrinkling of the flame is captured
directly.  By virtue of using a subgrid scale model to account for
unresolved turbulence, we expect that combined effects of unresolved
and resolved turbulence should result in the same global turbulent flame
speed independent of the resolution chosen.  Note that strict convergence,
for which an order can be defined, is not expected.  Higher resolutions
have both a smaller dissipation scale as well as a thinner reaction front,
thus posing different physical problems.  Instead we look for
consistency among resolutions.

\begin{figure}[tb]
\centering
\includegraphics[width=\columnwidth]{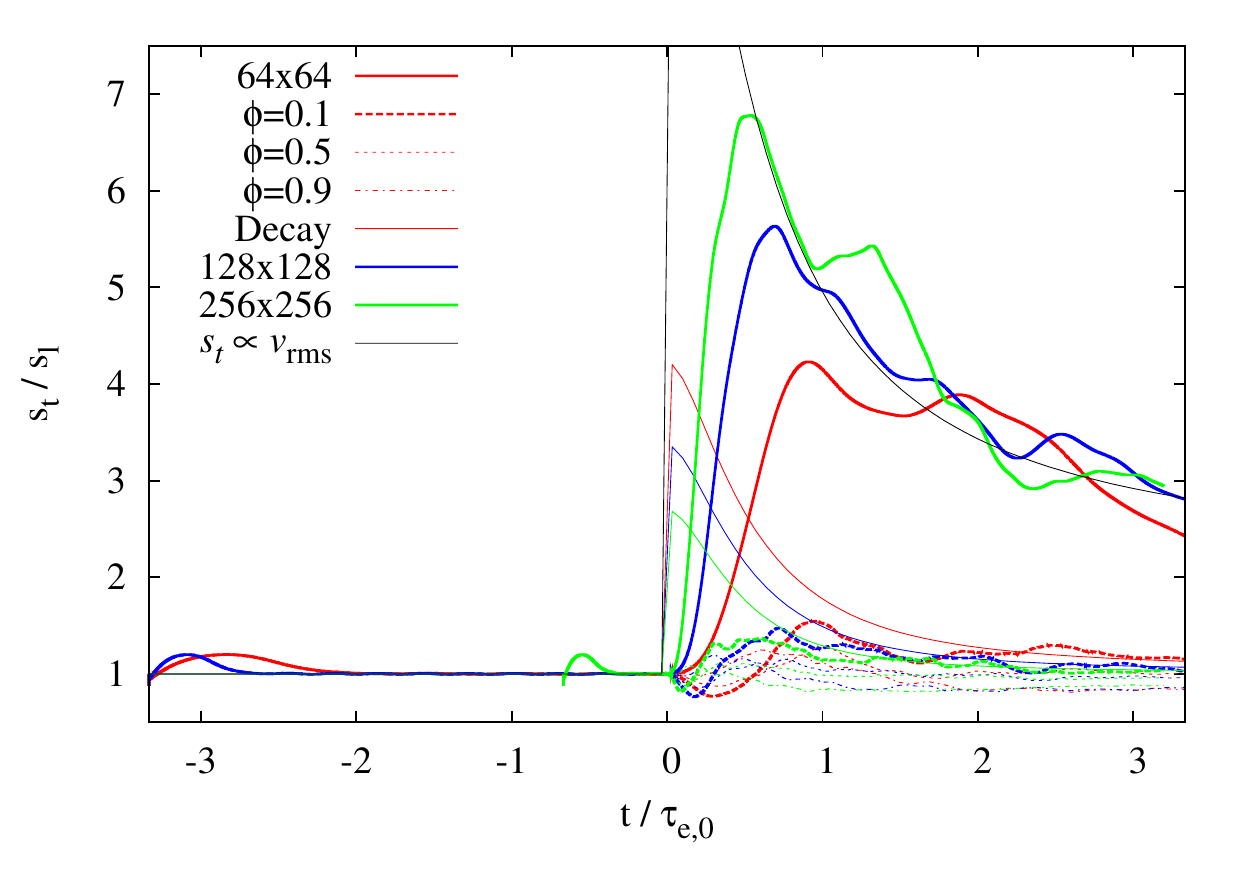}
\caption{
The ratio of the front propagation speed to input speed of the 
iso-contours of the reaction progress
variable for $\phi = \{0.1, 0.5, 0.9\}$ (dashed, dotted, and dot-dashed
lines, respectively) as well as the turbulent flame speed (thick solid
lines) are compared to the estimated turbulent flame speed from the
decaying turbulence (thin solid lines). The thin black line provides
the Damk\"ohler scaling relation for decaying turbulence.
Cross-sectional resolutions of 
64x64 (red), 128x128 (blue), and 256x256 (green) are compared by 
super-imposing a turbulent
velocity field in the fuel with $v_{\rm rms}=\unit{100}{\kms}$. The flow
is not
driven, so the turbulence field decays from numerical dissipation at the
grid scale and the estimated front-propagation speeds of iso-contours
of $\phi$ return to the input front-propagation speed. An SGS model is
not used for these calculations, and the solution does not converge with
resolution.
\label{fig:notfi}}

\end{figure}

\begin{figure}[tb]
\centering
\includegraphics[width=\columnwidth]{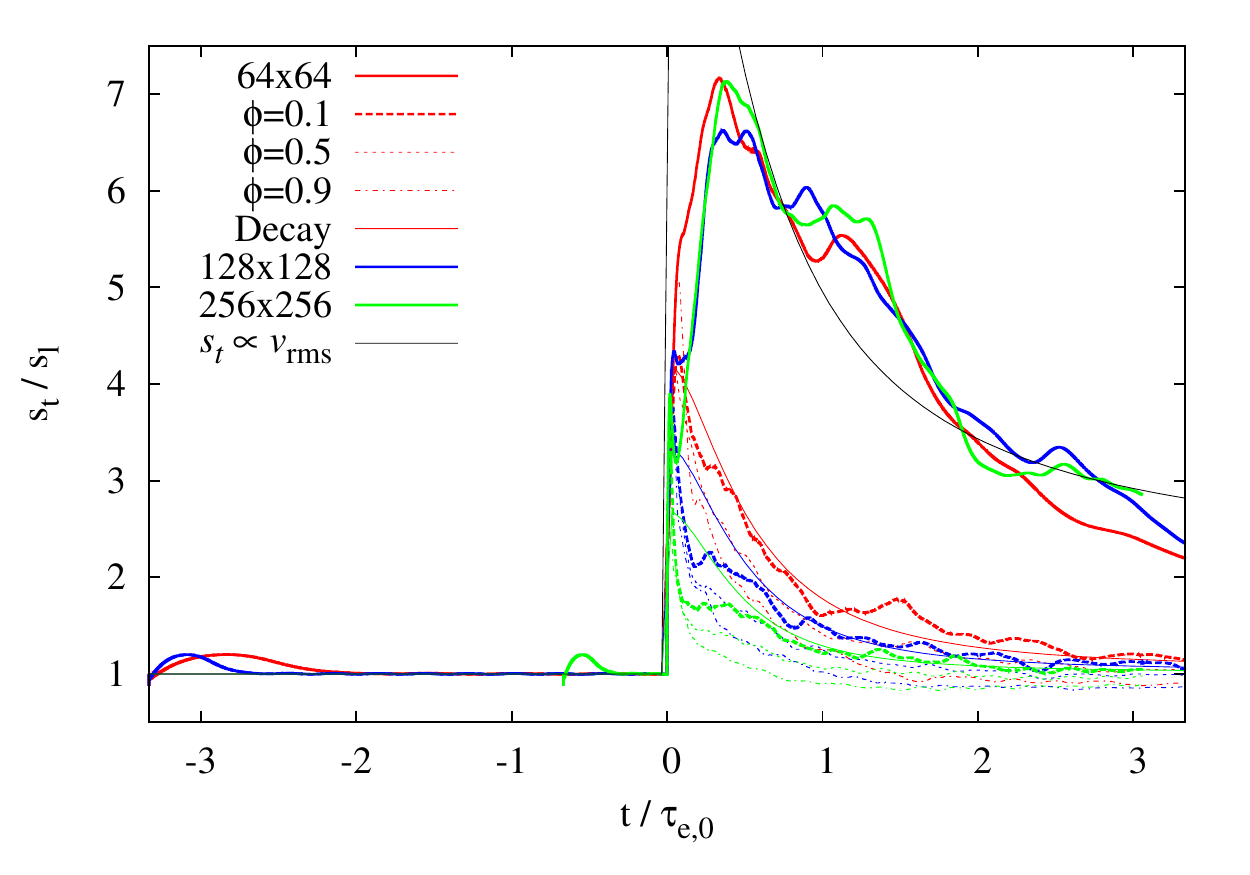}
\caption{
The same as \figref{fig:notfi}, except that the power-law flame wrinkling 
model is used and convergence with resolution is achieved.
\label{fig:turbulent}}
\end{figure}

\figref{fig:notfi} and
\figref{fig:turbulent}
show the results of simulations with $64^2$, $128^2$,
and $256^2$ cells resolving the plane orthogonal to the direction of
propagation. \figref{fig:notfi} shows results if a TFI model is not used,
and \figref{fig:turbulent} shows results using the TFI model implemented
in this work. Initially, the turbulent velocity field is scaled to $v_{\rm
rms} = \unit{100}{\kms}$, giving a Gibson scale of $\unit{500}{\cm}$, or
1/3000 of $L$.
The global turbulent flame speeds computed from the
fuel consumption rates are given by the thick solid lines, which are
consistent with one another when the TFI model is included.
When no TFI model is included, the burning rates obtained depend strongly on
resolution.  If the burning rates with the TFI model are correct, then the 256
resolution case without the TFI model is near convergence for the gross burning
rate.  Interpreting this result naively, neglecting subgrid TFI is not too
bad when $\Delta_x \lesssim 10\lambda_G$.  By contrast even the 64 cell
resolution case with the TFI model included is already converged.  This is
consistent with the idea that the TFI model should give accurate results
regardless of $\Delta_x/\lambda_G$ as long as there is enough resolution
for a turbulence cascade.

The front-propagation speed based on areas of iso-surfaces
of the reaction progress variable $\phi=\{0.1, 0.5, 0.9\}$ are given as
dashed, dotted, and dot-dashed lines with the $\phi=0.1$ line thickened
to illustrate that with increased resolution, the unresolved portion of
the inertial range contributes less to the global turbulent flame speed.
From the case without a TFI model, it appears that the disruption of the
reaction-diffusion front, which might appear as significant differences
among the iso-surface areas, does not occur even though the lower
resolution cases are clearly under-representing the total flame
propagation.  This is likely due to the proximity of the dissipation scale
to the flame width, as both are related to the grid resolution.

Because the turbulence field is not driven in the fuel ahead of the
flame, the turbulent velocity decays relatively quickly.
In a separate
calculation, the $256^3$-resolution run from \secref{sec:turbulence}
was restarted with no stirring to characterize the decaying turbulence
field. The decaying power spectra maintain a Kolmogorov spectrum, while
$v_{\rm rms}$ is described by
\begin{equation}\label{eq:decay}
v_{\rm rms,decay}(t) = e^{-t/\tau_e(t)} {\rm ,}
\end{equation}
where $\tau_e(t) = A*t + B + C/t$ is the eddy turn-over timescale. 
As the velocity decays, the eddy turn-over timescale increases. 
The constants $A$, $B$, and $C$ were determined using
least-squares fits with $A = 0.4353 \tau_{e,0}$, $B = 0.4454 \tau_{e,0}$,
and $C = 0.1973 \tau_{e,0}$. This fit is consistent with the well-known
result that $\left<u'^2\right> \propto t^{-1.3}$ for decaying turbulence
after $\approx \tau_{e,0}$, but also accurately captures the decay
for $t < \tau_{e,0}$. An eddy turn-over time is approximated
by $\tau_{e,0} = L / v_{\rm rms}$ where $L = \unit{15}{\km}$. Therefore, an
estimate of the turbulent velocity at a particular scale is provided by
\begin{equation}\label{eq:velscale}
\turbD(t) = v_{\rm rms,decay}(t) \left(\frac{\Delta}{L}\right)^{1/3}
{\rm .}
\end{equation}
An estimated turbulent flame speed at scale $\Delta$ is calculated
analytically using the TFI model and provided 
for comparison in \figref{fig:notfi} and \figref{fig:turbulent} as thin lines
colored corresponding to the simulation. The expected result
from Damk\"ohler scaling ($s_t \propto u'$) at scale $L$ is given
by the thin black line in each figure. This estimate reasonably
predicts the behavior of the flame speed after some delay time that
appears to be related to $\tau_{\rm flame}$. The reaction-diffusion front
requires $1-2 \tau_{\rm flame}$ to equilibrate to the flow around it.
Without the utilization of a TFI model, $\tau_{\rm flame}$ is
resolution-dependent (see \figref{fig:notfi}). When a TFI model is used,
the front propagation speed is increased such that $\tau_{\rm flame}$
is resolution-{\emph independent}, resulting in the better convergence
seen in \figref{fig:turbulent}. 


While we have shown convergence over a factor of $4$ in resolution,
this result does not preclude the possibility that the TFI model
may not accurately represent the TFI over a broader range of scales
such as those being modeled in \SNIa simulations. To extend our
analysis to larger computational domains is prohibitively expensive, and
we leave this concern to future work.

\section{Future Considerations}
\label{sec:future}

In order to improve upon this model, DNS calculations of single
vortex-flame interactions should be computed for low-\PrN,
high-\Le flames to verify the efficiency function
for astrophysical flames. Unfortunately, these calculations are
expensive and outside the scope of the present study. \citet{Zingetal01}
provided preliminary results of exactly this calculation; however, a more
comprehensive and complete analysis is required.

Future calculations should also consider flame curvature and stretch
effects, especially for high-\Le flames. Currently, this model
assumes \citetalias[\eq{6}]{Charetal02a},
\begin{equation}\label{eq:CMV6}
- \left< w \divr \bvec{n} \right>_s \approx
  \slamO \left| \left< \divr \bvec{n} \right>_s \right|
{\rm ,}
\end{equation}
when $\eta_c>\dlamO$,
to describe the destruction of flame surface density due to the
laminar flame propagation. Here, $w$ is the flame front displacement
speed, $\left|\left<\divr \bvec{n} \right>_s \right|$
is the mean curvature of the flame, and \slamO is the unstrained
laminar flame speed. However, if one considers thermodiffusive effects
for flames with arbitrary $\Le$, then
\begin{equation}
-w = \slamO \left[ 1 + \left(\frac{1}{\Le} - 1\right) \frac{\Ze}{2} \divr \bvec{n} \right]
{\rm ,}
\end{equation}
to linear order in curvature~\citep[see, \eg,][]{Dursetal03}, where \Ze is the
Zeldovich number that describes the strength of the temperature sensitivity
of the driving reaction. For $\Le \approx 1$,
\eqref{eq:CMV6} is appropriate; however, for $\Le \ne 1$, as in thermonuclear
flames, it is not a good approximation.

Still, one could assume that regions of positive curvature never completely quench
the flame (\ie, $w > 0$ always), and
\begin{equation}
-\left< w \divr \bvec{n} \right>_s \approx \slamO \left[\left| \left< \divr \bvec{n} \right>_s \right|
+ \left(\frac{1}{\Le} -1\right) \frac{Ze}{2} \left< \left(\divr \bvec{n} \right)^2 \right>_s \right]
{\rm .}
\end{equation}
In this case, it is possible for the mean curvature to be zero, yet for the average front
propagation speed to be significantly different from $\slamO$. In this case,
the inner cutoff scale $\eta_c$ that describes a fractal-like flame would be better
defined by the root-mean-squared flame surface curvature. Of course, the relation
between $\left< \divr \bvec{n} \right>_s$ and $\left< \left(\divr \bvec{n}\right)^2\right>_s$
would still need to be established, and would likely depend on the net strain rate and/or
the Damk\"ohler number to indicate how easily the flame is packed by turbulence.

While the inverse mean curvature of the flame is prevented from
becoming smaller than the laminar flame width artificially, more
physical constraints may be constructed by considering flame merging
and quenching processes. As the mean curvature increases, higher-order
flame surface density destruction terms due to merging may become
important~\citep{MenePoin91}. The inclusion of merging and quenching
processes into the model may naturally prevent the mean curvature of the
flame from becoming too large and will provide a smooth transition from
moderate mean curvature to strong mean curvature. Recent calculations
by \citet{PoluOran10,PoluOran11} indicate that the transition to
``distributed burning'' or ``broken reaction zones'' require higher
turbulent intensities than previously thought. This result implies that
the flame remains a well-defined concept into the ``thin reaction zone''
and that model improvements may extend the range of validity to
higher-intensity turbulence.

As a possibly essential addition, the effect of unresolved RT modes
should be self-consistently incorporated into the subgrid scale
TFI model. This instability could be incorporated into
the energy function $E(k)$ using an appropriate efficiency function with
a critical scale being the so-called ``fire-polishing'' scale $\ell_{\rm
fp}$.  \citet{Ciaretal09} has shown that the turbulent energy spectrum for
$\bvec{k} \parallel \bvec{g}$ follows that expected from RT
scaling, while for $\bvec{k} \perp \bvec{g}$ the energy spectrum follows
homogeneous and isotropic turbulence. We plan to include the effect of
increased flame surface due to RT instability in future modeling efforts. 
Currently, we instead choose the turbulent flame speed based on the 
dominant effect such that
\begin{equation}\label{eq:choose}
\sturb = {\rm max}\left( s_{\rm TFI}, s_{\rm RT} \right)
{\rm ,}
\end{equation}
where $s_{\rm TFI}$ is the turbulent flame speed estimated from the
TFI model described in this work and $s_{\rm RT}$ is the turbulent
flame speed estimated from RT instability described in \citet{Khok95} and
\citet{Townetal07}. However, we still prevent \sturb from increasing
beyond the limit $\slamO (1 + \Delta/\dlamO)^\gamma$, which provides a
natural quenching process as burning progresses toward lower densities
(see \figref{fig:colin_st}).

\section{Comparing TFI Models in Simulations of \SNeIa}
\label{sec:sims}

We perform three full-star, 3D simulations of the deflagration
phase of a centrally-ignited supernova using \flash with the flame
resolved at $4\km$ and the star at $16\km$.
One simulation is performed
without an explicit TFI model, which only accounts buoyancy effects by
setting the turbulent flame speed to
\begin{equation}\label{eq:st_buoy}
\sturbD = s_{\rm RT} = 0.5 \sqrt{Agm\Delta} {\rm ,}
\end{equation}
where the Atwood number
\begin{equation}\label{eq:atwood}
A = \frac{\rho_{\rm f} - \rho_{\rm a}}{\rho_{\rm f} + \rho_{\rm a}}
\end{equation}
describes the density change across the flame,
$g$ is the local gravitational acceleration, and $m$ is a
calibrated constant determined to be $m \sim 0.04-0.06$~\citep{Townetal08}.
We will call this implicit TFI model the buoyancy-compensation  model.
Two more simulations compare the relation between $\turbD$ and $\sturbD$.
One assumes that the TFI is scale invariant which follows the 
simple prescription described by \citet{Poch94} and 
utilized by \cite{Schmetal06a} given as
\begin{equation}\label{eq:pocheq}
\sturbD = \sqrt{{\slamO}^2 + C_t{\turbD}^2} {\rm ,}
\end{equation}
where $C_t$ is a constant taken to be $4/3$, the same as that
used in \citet{Schmetal06b}. We will call this model the scale invariant
TFI model. The final simulation employs the power-law flame wrinkling 
model described in this work.
Both of the latter simulations utilize the measure of turbulence described
in \secref{sec:turbulence}.
Note that the power-law flame wrinkling method was developed by CMV to not
require any assumptions about integral scale quantities.  The modeling is
entirely based on measurement of turbulence characteristics on scales just
above the grid scale.

All other aspects of the simulations are fixed relative to each other. We 
do not include an initial background turbulent
velocity field as expected from \citet{Zingetal09}; therefore, these tests
serve to highlight the minimum difference expected from the choice of TFI
model. We ignite a near-Chandrasekhar mass WD with a central density $\rho_c
= 2.2\times10^9 \grampercc$ and central temperature $T_c = 7\times10^8 K$.
The initial WD is described in \citet{Jacketal10} as having an isentropic, 
carbon-depleted core with
$X(\carbon) = 0.4$ and an isothermal outer layer with $X(\carbon) = 0.5$,
where $X(\carbon)$ is the carbon mass fraction. We initialize the simulation
by placing 844 spherical hot spots with radius $r = 10\km$ randomly 
within the inner $150\km$ of the star. We follow the evolution until the
flame reaches a density of $\sim 10^7 \grampercc$, at which point a
detonation is expected in the deflagration-to-detonation paradigm.

As \flash heavily utilizes adaptive mesh refinement capabilities, we perform
two test simulations to quantify the effect the refinement criteria has on
the development of turbulence. 
If burning generates significant turbulence ahead of the flame,
then those regions should be resolved to the same scale as the flame to
allow turbulence to develop without artificial numerical dissipation.
One simulation is conducted with the inner $1024\km$ uniformly refined to
capture any turbulence generated that may influence the flame propagation,
while the other only refines energy-generating regions. Differences in the
evolution of these two simulations are negligible, and we conclude that a
uniformly refined region is not necessary for a non-turbulent initial
condition.

While end-to-end simulations were not performed, qualitative differences may
be discerned from the deflagration phases simulated. The largest differences
occur between the explicit and implicit TFI treatments, with the implicit,
buoyancy-driven treatment burning less fuel overall, and therefore expanding
less rapidly. In addition, the simulation with implicit TFI produced fewer
stable Fe-group elements. We conclude from these results that while using
the implicit TFI model causes the star to expand less, the burning rate is
even smaller, resulting in fewer stable Fe-group elements.

\begin{figure}[tb]
\centering
\includegraphics[width=\columnwidth]{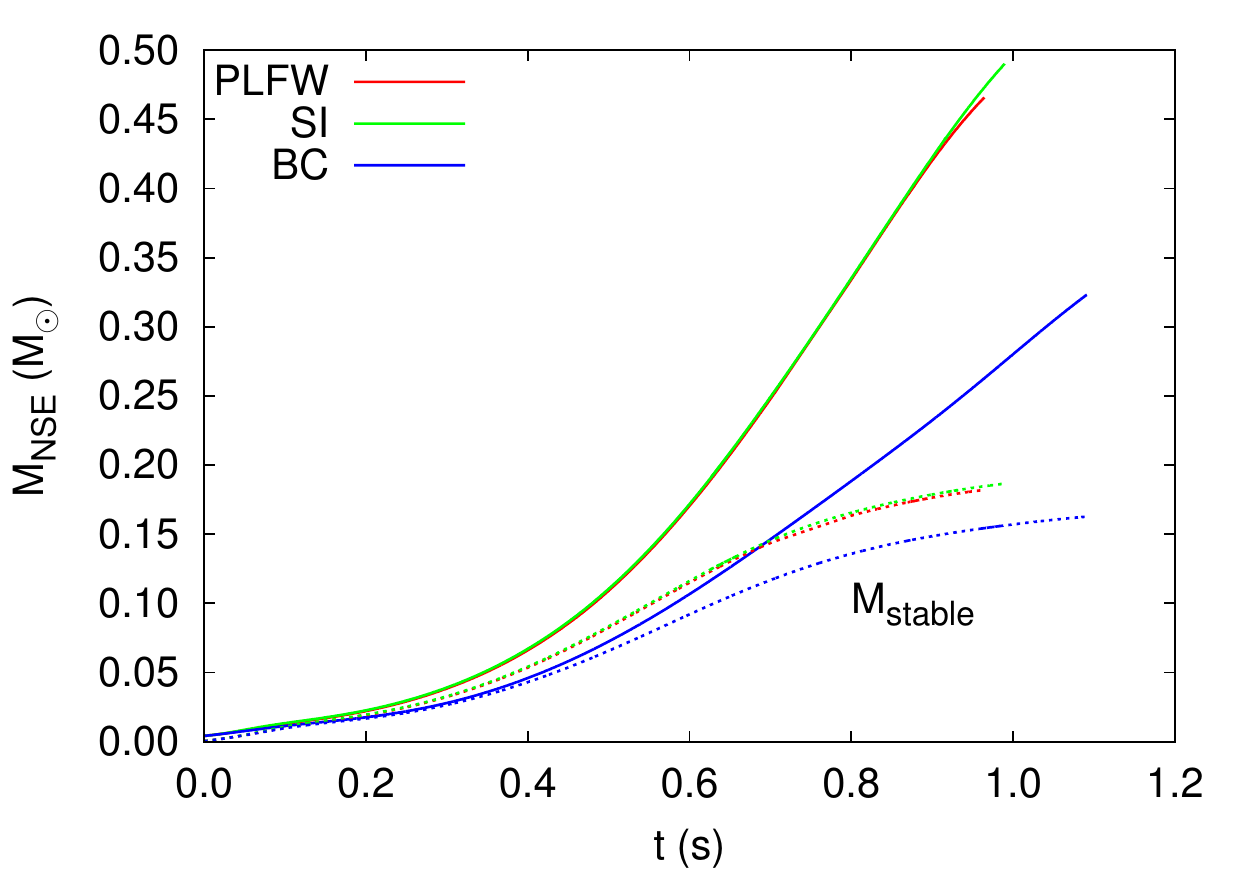}
\caption{
The yield of material in nuclear statistical equilibrium (NSE; solid lines)
and an estimate of the yield of stable Fe-group elements (dotted lines) as
a function of time for the power-law flame wrinkling (red), scale invariant 
(green), and buoyancy-compensation (blue) simulations. The simulations with 
explicit TFI 
(power-law flame wrinkling and scale invariant) show a larger
production of material in NSE as well as stable isotopes during the
deflagration phase; albeit, the difference in stable isotope production is
smaller.
\label{fig:ige}}
\end{figure}

\begin{figure}[tb]
\centering
\includegraphics[width=\columnwidth]{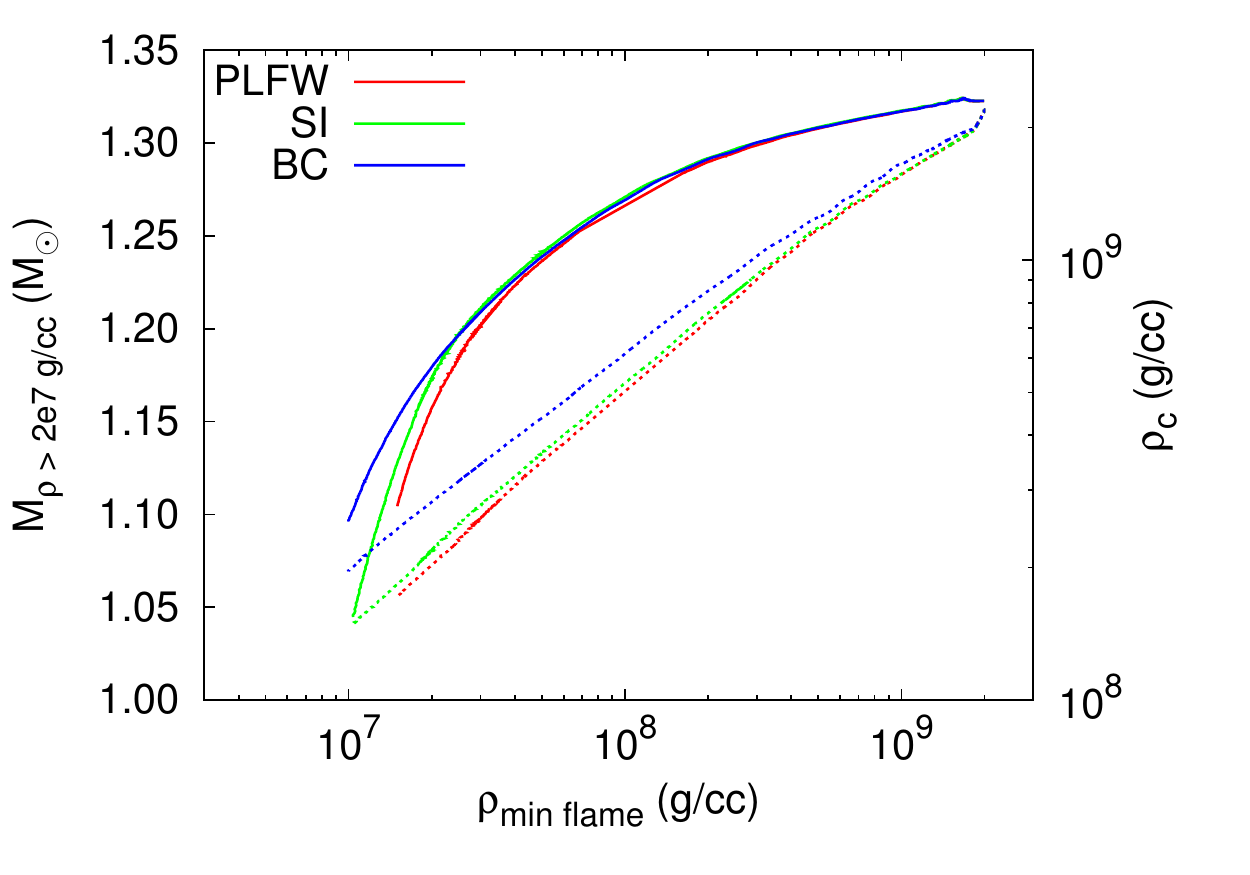}
\caption{
The amount of material at high density (solid lines; left y-axis) and 
central density (dotted lines; right y-axis) as a function of the minimum
flame density for the power-law flame wrinkling (red), scale invariant 
(green), and buoyancy-compensation (blue) simulations.
The material with $\rho > 2\times10^7 \grampercc$ is representative of the
yield of Fe-group elements if a detonation were to be initiated at a
particular instant in time (or a particular flame density).
The simulations with explicit TFI (power-law flame wrinkling and scale 
invariant) show a faster expansion rate as evidenced by a
steeper slope in the central density over
the simulation with implicit TFI (buoyancy-compensation). This translates 
to nonlinear behavior
in the estimate of final Fe-group yields with differences in the final
yields diverging faster than the central densities as a function of the
minimum flame density.
\label{fig:expanrise}}
\end{figure}

\begin{figure}[tb]
\centering
\includegraphics[width=\columnwidth]{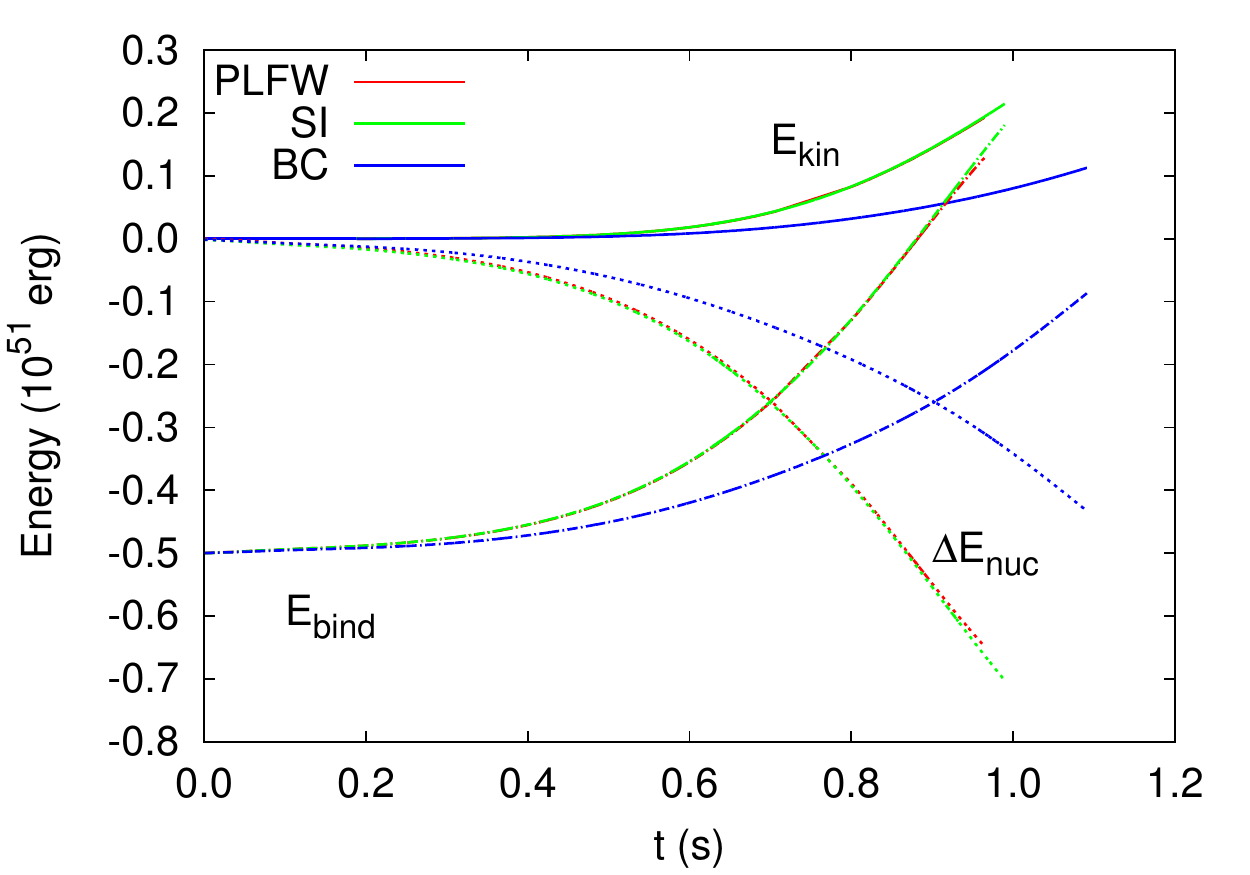}
\caption{
Energy budget as a function of time for the power-law flame wrinkling (red), 
scale invariant (green), and buoyancy-compensation (blue) simulations. The 
kinetic (solid lines), change 
in nuclear (dotted lines), and binding (dot-dashed lines) energies are 
compared. Nuclear energy is released as fuel with high rest-mass energy 
is processed to ash with
low rest-mass energy. The released energy is converted into internal and
kinetic energy, and serves to gravitationally unbind the star.
\label{fig:ecompare}}
\end{figure}

Figures
\ref{fig:ige} through \ref{fig:ecompare} support this conclusion.
\figref{fig:ige} compares the amount of material in nuclear statistical
equilibrium (NSE) as well as the portion estimated to end up in the form of
stable isotopes as a function of simulation time (see \citet{Jacketal10} for
how this is estimated from the local $Y_e$). The explicit TFI models
enhance burning in the turbulent regions dominated by the KH
shear instability along the sides of the rising RT unstable
plumes. This enhanced burning results in $\approx 0.2 M_\odot$ more material 
being processed to NSE as compared to the implicit TFI model. The difference
in the amount of stable isotopes is less pronounced with a difference of
$\approx 0.05 M_\odot$ between the models. While the final yield of Fe-group
material is difficult to estimate from the deflagration phase alone, the
final yield of stable Fe-group material should not differ very much from
that produced during the deflagration. The separation in the curves that
show the total NSE material and the stable portion occur at different times
with the explicit TFI models separating around $0.2 \second$ and the
implicit TFI model around $0.4 \second$. While the implicit TFI model burns
material at a slower rate, the expansion rate is also slower, allowing the
processing of NSE material to occur longer at high density where weak
processes are important. We expect the difference between explicit and
implicit TFI models to be more stark with higher central density progenitor
WDs. \figref{fig:expanrise} compares the expansion history of the WD in
terms of the progression of the flame in density-space. We find this
presentation most useful in assessing the outcome of a simulation in the
context of the DDT paradigm as DDT is expected to occur when the flame
reaches $\sim 10^7 \grampercc$. We plot two measures of the expansion as a
function of the minimum flame density: the amount of material with a density
greater $2\times10^7 \grampercc$ on the left y-axis and the central density on
the right y-axis. From previous studies~\citep{Townetal09}, we find that
the amount of material with with $\rho > 2\times10^7 \grampercc$ reasonably
predicts the yield of NSE material if a detonation were to be triggered at
that instant in time. From \figref{fig:expanrise}, we expect $\sim 0.1
M_\odot$ difference in NSE yields between the buoyancy-compensation TFI model and the scale invariant TFI
model, if DDT
were to occur at $\rho_{\rm DDT} = 10^7 \grampercc$. If DDT is delayed, the
difference is expected to grow substantially since the expansion rates are very different. This figure demonstrates the only clear deviation
between the two explicit TFI treatments, the power-law flame wrinkling 
and scale invariant models. The power-law flame wrinkling
model appears to expand slightly more for the same flame progression that
may result in a difference in the NSE yield of $\sim 0.05-0.1 M_\odot$ for
$\rho_{\rm DDT} = 10^7 \grampercc$, although if DDT were delayed the
difference could be larger. Finally, \figref{fig:ecompare} shows the energy
budget for the three different simulations, which supports the previously
drawn conclusions. The implicit TFI model burns material at a slower rate,
and releases less nuclear binding energy. This serves to delay the unbinding
of the WD and results in a lower kinetic energy as compared to the explicit
TFI model simulations. Additionally, note that both simulations with
explicit TFI show the star unbinds during the deflagration phase, consistent
with the centrally-ignited, pure deflagration results of \citet{ropke_2007_ac}.

As simulations were only performed of the deflagration phase, only
qualitative estimates of the final explosion outcome are provided. In
general, it is clear that turbulence driven by KH instabilities
in the wake of rising, RT unstable plumes serves to enhance
burning in those regions than would otherwise be predicted by only
considering buoyancy instabilities. The resulting increase in the burning
rate leads to a different evolution of the flame and
expansion of the star. We estimate that enhanced burning due to an explicit
TFI treatment will result in $\gtrsim 10\%$ decrease in the total Fe-group
yield, as well as $\gtrsim 30\%$ increase in the yield of stable Fe-group
elements.
In addition, the decrease in the efficiency of turbulence to wrinkle the
flame at low density for the power-law flame wrinkling model 
compared to the scale invariant model may
contribute to an additional $\sim 10\%$ decrease in the total Fe-group
yield. As a final note, the differences highlighted between the two 
explicit TFI models, the power-law flame wrinkling and scale invariant 
models, only highlight differences in
how the turbulence measure is {\it used}, not how it is obtained.
Simulations performed with the TFI model described by
\citetalias{Schmetal06b} may show greater differences owing to the
gradient-diffusion approximation for subgrid scale turbulent transport.

\section{Conclusions}
\label{sec:conclusions}

We have implemented a power-law flame wrinkling TFI model based 
on \citetalias{Colietal00}
and \citetalias{Charetal02a}. We calibrated a method used in \flash to
measure the turbulent velocity at the grid scale and have shown its
validity over a range of calculations with increasing resolution. We
compared and explored different implementation choices relevant for
astrophysical flames ultimately deciding that a low-\PrN extension
of the \citetalias{Charetal02a} TFI model is most appropriate for \SNeIa
simulation. We provided two convincing test problems of the TFI
model in \flash for laminar and turbulent flows. Of most importance,
we have shown that the global turbulent flame speed is consistent with
increasing resolution where more of the inertial range is resolved on
the computational domain. While this doesn't conclusively show that the
model captures TFI for the full inertial range present in the explosion,
the combined effects of the subgrid scale TFI
model with the resolved wrinkling of the model flame produce a consistent
burning rate independent of resolution over a factor of $4$.
Additionally, we have shown that
a laminar flame is recovered in the limiting case of laminar flow.

In order to understand how a power-law flame wrinkling model may effect the
outcome of simulations
of \SNIa, we compared full-star, 3D simulations of an \SNIa
utilizing different treatments for the
under-resolved TFI including one that considered only the effects of
buoyancy on the turbulent flame as well as two others that additionally
considered the directly measured turbulence local to the flame. The two
latter simulations considered differences in the utilization of the same
turbulence measure. We compared differences in the
evolution of the deflagration phase of an \SNIa between the buoyancy-only
model and the models that considered directly measured local turbulence. We
demonstrated that the buoyancy-only model does not capture enhanced burning
due to turbulence driven by the KH shear instability along the
sides of rising, RT-unstable plumes. This deficiency in the
buoyancy-only model leads to a reduced stable Fe-group yield and a larger
overall Fe-group yield, which suggests an over-production of radioactive
${}^{56}\mathrm{Ni}$ by $\gtrsim 0.2 M_\odot$ assuming DDT takes place when
the flame reaches $\rho \approx 10^7 \grampercc$. If DDT occurs at lower
densities, the difference is more pronounced. These estimates represent
a lower-bound as the simulations performed did not include a turbulent
background as expected from the slow smoldering of carbon during the process
of thermonuclear runaway~\citep{Zingetal09}. The turbulent background would
only serve to enhance differences observed.

The demonstration of scale convergence of the subgrid scale TFI model
for homogeneous, isotropic, decaying turbulence will enable resolution 
studies of \SNeIa simulations to determine whether
the assumption of these turbulence properties on
unresolved scales is valid. Failure to achieve convergence of the consumption
rate in \SNeIa with resolution will indicate that additional unresolved physics 
\emph{important to the explosion process} must be modeled. Such 
processes may be unresolved RT-unstable modes that contribute to flame
surface growth or non-equilibrium turbulence.

This model limits the growth of flame surface based on the fact
that the inverse mean curvature of the flame should not become smaller
than the laminar flame width. This condition was shown to be similar
to arguments that the turbulent flame should quench when $\gibson < \dlamO$,
although derived from the efficiency function describing the strain rate
of individual eddies within the cascade. While this effect does not appear
to be completely captured directly by the \citetalias{Colietal00} DNS calculations
of flame--vortex interactions, it could be verified by future calculations.

Future calculations of \SNeIa will also benefit from the reliable measure
of the turbulent velocity on unresolved scales in determining more
realistic conditions for DDT. Our recent calculations
\citep{Townetal09,Jacketal10,Krueetal10} determined DDT 
conditions simply by the requirement that the flame reach a particular
density, $\rho_{\rm DDT}$. This condition resulted in detonation ignitions
at the ``tops'' of rising plumes where turbulence is expected to be
relatively weak.  A more appropriate DDT condition, viable in 3D
simulations, is one based on the local turbulence intensity
\citep[e.g.][]{GoloNiem05}.  With such a more realistic DDT condition
based on $\turbD$, detonations will likely ignite in the turbulent regions
underneath plume caps.

Ideally, the TFI model would predict the DDT time and location during the
explosion, thus removing a free parameter from the model. This would allow
the explosion outcome to depend solely on the initial conditions of
the WD and the distribution of the first flames. The accurate prediction of DDT
is necessary to understand the range of properties of progenitor WDs that lead
to realistic explosions, and how
variations in those properties translate to variations in the explosion outcome.

\acknowledgements

The authors thank Alexei Poludnenko and Elaine Oran for useful discussions
that contributed to this work.  We also thank the anonymous referee for 
a careful
reading of the manuscript and many suggestions to improve its accessibility.
This work was supported in part by an award to APJ. from the National
Research Council Research Associateship Program at the Naval Research
Laboratory.
This work was supported by NASA through grant NNX09AD19G.
ACC also acknowledges support from the 
Department of Energy under grant DE-FG02-87ER40317. DMT received support
from the Bart J. Bok fellowship at the University of Arizona for part of 
this work. The authors acknowledge 
the hospitality of the Kavli Institute for Theoretical Physics, which is 
supported by the NSF under grant PHY05-51164, during the programs ``Accretion 
and Explosion: the Astrophysics of Degenerate Stars'' and ``Stellar Death and
Supernovae.''  The software used in 
this work was in part developed by the DOE-supported ASC/Alliances Center 
for Astrophysical Thermonuclear Flashes at the University of Chicago. We 
thank Nathan Hearn for making his QuickFlash analysis tools publicly 
available at http://quickflash.sourceforge.net. This work
utilized resources at 
the New York Center for Computational Sciences at Stony Brook 
University/Brookhaven National Laboratory which is supported by the U.S. 
Department of Energy under Contract No. DE-AC02-98CH10886 and by the State of 
New York.  Some simulations presented in this work were run on the Ranger
supercomputer at the Texas Advanced Computing Center as part of the Extreme
Science and Engineering Discovery Environment (XSEDE, formally TeraGrid),
which is supported by National Science Foundation grant number OCI-1053575.
Computations were also performed under the Department of Energy 2010 INCITE
allocation, {\em Fundamental Research in Type Ia Supernovae.}  

\bibliographystyle{apj}
\bibliography{shorttitles,JabRef_master}

\end{document}